# Early Australian Optical and Radio Observations of Centaurus A

*Peter Robertson*[A, C], *Glen Cozens*[A], *Wayne Orchiston*[A], *Bruce Slee*[A, B] *and Harry Wendt*[A]

[A] Centre for Astronomy, James Cook University, Townsville, QLD 4811, Australia

[B] Australia Telescope National Facility, CSIRO, PO Box 76, Epping, NSW 2121, Australia

[C] Corresponding author. Email: peter.robertson@jcu.edu.au

**Abstract:** The discoveries of the radio source Centaurus A and its optical counterpart NGC 5128 were important landmarks in the history of Australian astronomy. NGC 5128 was first observed in August 1826 by James Dunlop during a survey of southern objects at the Parramatta Observatory, west of the settlement at Sydney Cove. The observatory had been founded a few years earlier by Thomas Brisbane, the new governor of the British colony of New South Wales. Just over 120 years later, John Bolton, Gordon Stanley and Bruce Slee discovered the radio source Centaurus A at the Dover Heights field station in Sydney, operated by CSIRO's Radiophysics Laboratory (the forerunner of the Australia Telescope National Facility). This paper will describe this early historical work and summarise further studies of Centaurus A by other Radiophysics groups up to 1960.

## 1 Introduction

The radio source Centaurus A and its optical counterpart NGC 5128 comprise possibly the most important astronomical object in the history of Australian astronomy. Observations of the southern sky in Australia are as old as human settlement. There is a growing body of evidence that the importance of astronomical knowledge to Aboriginal culture was far greater than previously thought (see e.g. Norris 2008). With the arrival of British settlement in 1788, a rudimentary observatory was established at Sydney Cove in the new colony of New South Wales, but it was used primarily for time-keeping and as a navigational aid and did not produce any astronomical observations of significance (see Haynes et al. 1994).

Astronomy as a science in Australia began with the establishment of the Parramatta Observatory in 1822 by Sir Thomas Brisbane, the sixth governor of the British colony of New South Wales. Brisbane is best remembered today by the name of Australia's third largest city, the capital of the state of Queensland. The observatory carried out the first systematic survey of southern stars and the catalogue was eventually published in 1835. In 1826 Brisbane's assistant, James Dunlop, carried out another survey of southern nebulae and star clusters. Among the over 600 objects catalogued was an object later named NGC 5128. As we discuss in Section 2, Dunlop's catalogue was published in England in 1828, forty years after the establishment of the new colony, and the first astronomical paper published from Australia.

In mid-1947, just over 120 years after Dunlop began his survey, observations of the southern sky of a radically different nature began at Dover Heights in Sydney, the site of a former wartime radar research station operated by the Radiophysics Laboratory. John Bolton, Gordon Stanley and Bruce Slee using a novel technique known as sea-interferometry were able to confirm earlier reports from England that there was strong radio emission from a concentrated area in constellation of Cygnus. As we see in Section 3, by the end of 1947 the group at Dover Heights had discovered several other point-like sources of radio emission, including one in the Centaurus constellation.

As the number of these new radio sources grew, the most important challenge was to determine the nature of these sources and to see whether they could be identified with objects known to astronomers. As we see in Section 4, Bolton and Stanley organised an expedition to New Zealand in mid-1948 where better observing conditions meant that the positions of the known radio sources could be measured with greater accuracy. In mid-1949, just two years after the initial Cygnus detection, the Dover Heights group published a paper with the new and more accurate positions. For three of the sources they offered tentative identifications with known optical objects, each of which turned out to be correct. The Taurus A source coincided with the Crab Nebula and Virgo A with M87. The third source, Centaurus A, coincided with NGC 5128, the nebula discovered by James Dunlop in 1826.

After 1949 the focus of the Dover Heights group was to build a series of increasingly sensitive radio telescopes to survey the sky. The number of known radio sources soon ran into the hundreds. In Section 5 we describe further study of Centaurus A at Dover Heights. Rather than a point-like source, this nearby galaxy was shown to have a very large angular size with a complex internal structure where the strength of the radio emission varies markedly over the area of the source.

With the closure of the Dover Heights field station in 1954, other groups within the Radiophysics Laboratory studied Centaurus A at a number of other field stations operating in the greater Sydney area. In Section 6 we describe this work up until 1960.

## 2  The Discovery of NGC 5128

### 2.1 Setting the Scene:  The Parramatta Observatory

The galaxy NGC 5128 was discovered by James Dunlop on 29 April 1826 at Parramatta, a small settlement approximately 20 km west of Sydney, in the British colony of New South Wales. (We use the contemporary spelling 'Parramatta' throughout this paper, rather than 'Paramatta' as it was then known.) Dunlop was born at Dalry, 35 km south-west of Glasgow, Scotland, on 31 October 1793. At age fourteen he moved to nearby Beith where he lived with his father's twin brother and worked in a thread factory. James was not well educated. According to one of his biographers, John Service (1890), '… he had been a short time at school in Dalry, and when he went to Beith, he attended a night-school in the Strand ... But, beyond these meagre opportunities for education, he received no scholastic training whatever'. Dunlop had a 'natural aptitude and love [for mechanics and] when he was seventeen years of age, he was constructing lathes and telescopes and casting reflectors for himself' (Service 1890).

In 1820 Dunlop was introduced to Sir Thomas Brisbane, a meeting that led to Dunlop accompanying Brisbane to Australia. Brisbane was also born in Scotland and educated at the University of Edinburgh. He then joined the British Army where he rose to the rank of Lieutenant-General (Heydon 1966; Saunders 2004). During the Peninsular war he commanded a brigade which was heavily engaged in battles in France. Brisbane was knighted in 1815 in recognition of his service. The Duke of Wellington recommended him to the Colonial Office for higher duties and, in November 1820, Brisbane was offered the post of governor of the colony of New South Wales (see Figure 1).

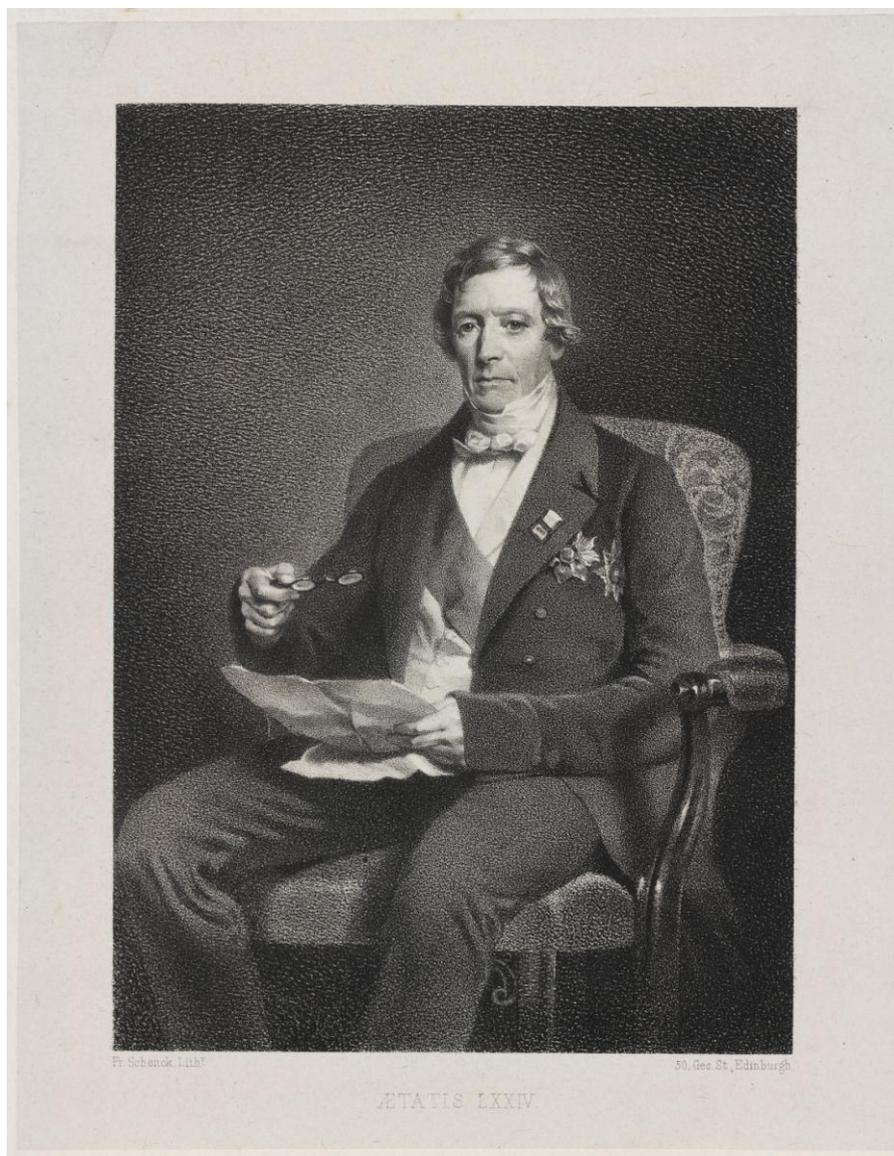

**Figure 1** - Sir Thomas Makdougall Brisbane (1773–1860) [portrait by F. Schenck, ca. 1850, courtesy State Library of NSW].

Aside from his military career, Brisbane was a fine example of a nineteenth century gentleman-scientist. He had a keen interest in astronomy and in 1808 built an observatory at his own expense in Ayrshire, south of Glasgow (Morrison-Low 2004). For the work there he was made a Fellow of the Astronomical Society of London. On his appointment as Governor of New South Wales, Brisbane decided he would build another observatory to explore the largely unchartered southern skies. He knew he would be too busy with his own duties as governor, so he hired two assistants – again at his own expense. One was James Dunlop for his talents as an instrument maker, the other was Christian Carl Ludwig Rümker, a mathematics teacher originally from Germany (Bergman 1967).

After a six-month voyage via Rio de Janeiro onboard the *Royal George*, Brisbane and his assistants arrived in Sydney in November 1821. Brisbane immediately arranged for the construction of the observatory next to Government House at Parramatta, west of the main settlement at Sydney Cove. Parramatta Observatory was completed within months (see Figure 2) and fitted out with Brisbane's instruments and personal library. One of the principal instruments was a 9.2 cm diameter transit telescope which would be used to survey the southern sky. Other notable instruments included a Reichenbach repeating circle and an 8.3 cm refracting telescope (see Lomb 2004).

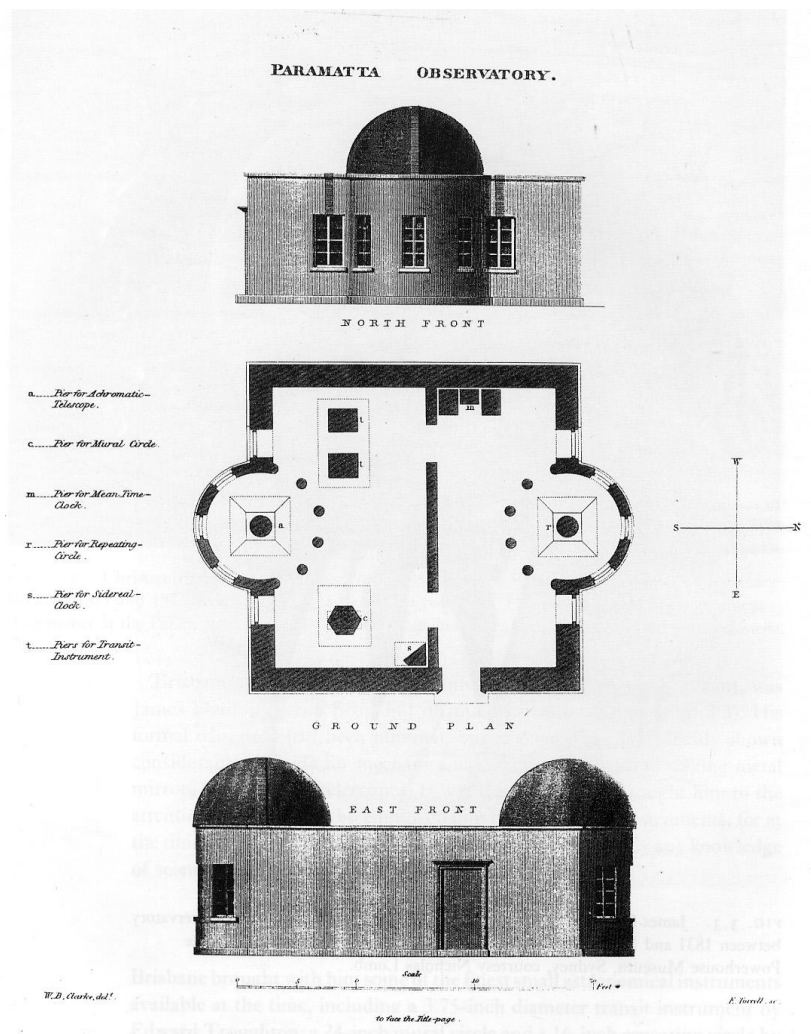

**Figure 2** - Plans for the Parramatta Observatory from a sketch by the Reverend W. B. Clarke in 1825.

The observatory got off to a flying start when in June 1822, only a month after observations began, Dunlop and Rümker observed the return of Encke's Comet. Dunlop found the comet at the position calculated by Rümker. After Halley's Comet in 1758 this was only the second occasion the return of a comet as predicted had been observed and it confirmed that comets, like planets, obeyed Newton's laws. The importance of the recovery of the comet can be gauged by the immediate award to Rümker of 100 pounds by the Astronomical Society of London (Bergman 1967).

Aside from Encke's Comet, Rümker and Dunlop embarked on an intense period of observing and cataloguing the stars of the southern sky. However, despite this very promising start, the productive years of the Parramatta Observatory were short-lived. After a series of disputes between Rümker and Brisbane, Rümker left Parramatta after just two years and established a farming property south-west of Sydney. By then Rümker had observed and recorded more than 2000 stars in preparation for a star catalogue. In Rümker's absence, Brisbane trained Dunlop to use the instruments and Dunlop finally completed the star catalogue observations in March 1826. Referred to as the *Parramatta Catalogue of Stars* or sometimes as the *Brisbane Catalogue*, the final version was eventually published in London in 1835 (Brisbane 1835). Although Dunlop and Rümker were mentioned in the preface, Brisbane was listed as the sole author, with William Richardson of the Royal Observatory at Greenwich acknowledged for the reduction of the data.

Like Rümker, Brisbane's stay at Parramatta was short-lived. Although, as governor, he introduced a number of notable reforms to the fast growing colony, Brisbane probably made a better astronomer than he did a politician. His critics prevailed and he was recalled to England by the Colonial Office at the end of 1825, just four years after his arrival. The colonial government agreed to purchase Brisbane's astronomical library and instruments in order to keep the Parramatta Observatory functioning.

*2.2 Dunlop's Catalogue of Nebulae and Star Clusters*

After Brisbane's departure, Dunlop also left the Observatory and moved to a cottage located approximately 1 km to the east (see Figures 3 and 4). In the backyard of his house he erected a nine-inch (23 cm) diameter reflecting telescope and in April 1826 began a punishing schedule of observations. Over the next seven months he catalogued the 629 objects that are listed in 'A Catalogue of Nebulae and Clusters of Stars in the Southern Hemisphere observed in New South Wales' (Dunlop 1828) and a further 253 double stars which appeared in 'Approximate Places of Double Stars in the Southern Hemisphere, observed at Parramatta in New South Wales' (Dunlop 1829). Dunlop worked on the double-star catalogue when the Moon was bright and on the non-stellar catalogue when it was dark. In the introduction to the first of the catalogues he noted:

> "The observations were made in the open air, with an excellent 9-feet (2.7 m) reflecting telescope, the clear aperture of the large mirror being nine inches (23 cm). This telescope was occasionally fitted up as a meridian telescope, with a strong iron axis firmly attached to the lower side of the tube nearly opposite the cell of the large mirror, and the ends of the axis rested in brass Y's, which were screwed to blocks of wood let into the ground about 18 inches, and projecting about 4 inches above the ground; one end of the axis carried a brass semicircle divided into half degrees and read off by a vernier to minutes. The position and index error of the instrument were ascertained by the passage of known stars. The eye end of the telescope was raised or lowered by a cord over a pulley attached to a strong wooden post let into the ground about two feet: with this apparatus I have observed a [north-south] sweep of eight or ten degrees in breadth with very little deviation of the instrument from the plane of the meridian, and the tremor was very little even with considerable magnifying power."

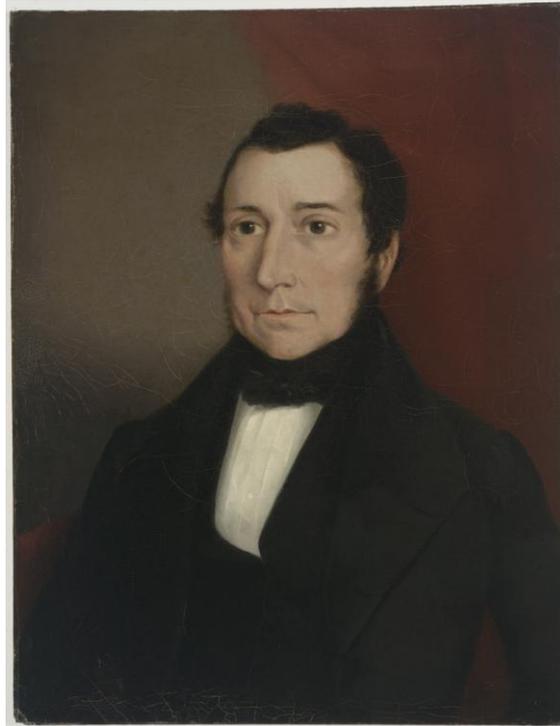

**Figure 3** - James Dunlop (1793–1848) [portrait by Joseph Backler, ca. 1843, courtesy State Library of NSW].

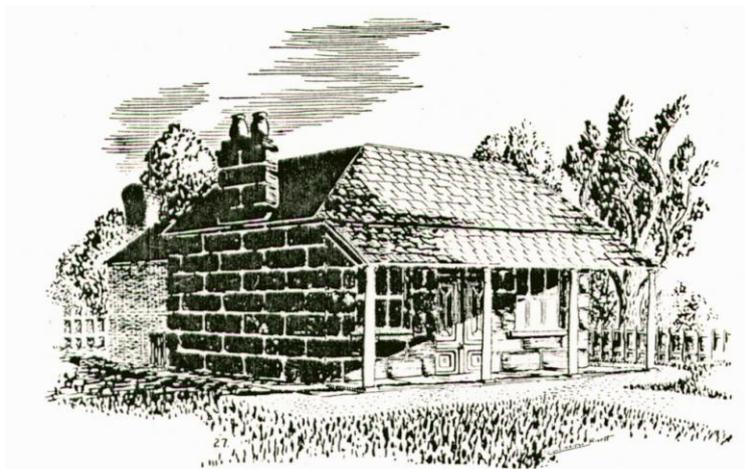

**Figure 4** - The stone cottage where James Dunlop lived in 1826 at the time of his discovery of NGC 5128, from a sketch by Collinridge Rivett (see Rivett 1988). Dunlop spent 1827–31 in Scotland before returning to Parramatta Observatory where he spent the remainder of his career.

Dunlop started his catalogue of nebulae and star clusters on 27 April 1826 when he observed the four open clusters NGC 3532, IC 2714, Melotte 105 and NGC 3766. Two nights later on Saturday 29 April he observed and recorded nine objects. The end of astronomical twilight occurred at 18:42 pm and the night was very productive, the objects seen (described by their contemporary names) being:

1. The open cluster NGC 3293, seen at 19:58. Discovered originally by Lacaille in 1752.
2. The open cluster NGC 3330, discovered by Dunlop at 20:00.
3. The open cluster D224=Harvard 6 was discovered at 21:56. Herschel missed this open cluster.
4. The globular cluster NGC 4833, seen at 22:16. Also discovered by Lacaille.
5. The galaxy NGC 4945, discovered at 22:23. This was Dunlop's first galaxy discovery.
6. The open cluster D244=Harvard 8, discovered at 22:35. Herschel missed this open cluster.
7. **The galaxy NGC 5128, discovered at 22:43.**
8. The galaxy M83, seen at 22:55. Also discovered by Lacaille in 1752.
9. The globular cluster NGC 5286, discovered at 23:03.

On that night, the 44% sunlit Moon rose at midnight and the magnitude 6.5 Comet Pons 1825 IV, which was near Psi Lupi, transited at 00:52.

Dunlop had a particularly busy hour from 21:56 to 22:55 when he observed six objects including four new ones. NGC 4945 and NGC 5128 were the first galaxies discovered south of declination $-33^{o}$ since the Magellanic Clouds. (In Sir William Herschel's earlier catalogue the most southerly galaxy was at declination $-32^{o}$ 49'.) NGC 5128 was recorded by Dunlop as number 482. His description and the diagram he drew are shown in Figures 5 and 6. His description of NGC 5128 as 'singular' probably means 'remarkable' or 'outstanding'.

| No. | Synon. | R A. 1830.0. h. m. s. d. | N P D. 1830.0. o  '  " | Description, Remarks, &c. | Sweep. |
|---|---|---|---|---|---|
| 3501 | Δ. 482 | 13  15  28.7 | 132  8  3 | [A nebula consisting of two lateral portions, and] no doubt of a small streak of nebula along the middle of the slit or interval between them, having a star at its extremity. See fig. 2, Pl. IV. Position of the slit, 124.7 ; of the star, with another ✳ near the nebula and south of it, 332.3 ; other stars also laid down (see description of the figure). A most superb calm night ; objects admirably defined. Shown to a bystander (J. R.) who saw it as figured and described. | 458 |
|  |  | 29.5 |  8  8 | [Two nebulæ, or two portions of one separated by a division or cut.] The cut is broad and sharp. The two nebulæ are very nearly alike. Perhaps the slit is larger towards the n p end, where there is a star between them. There is certainly a very feeble trace of nebula, an island as it were, running from this star between the sides of the slit. N.B. No "moon-light effect" seen between the edges. Night very fine. Pos of the slit, 120°.3. The place taken is that of the star within the slit. | 455 |
|  |  | 31.7 |  7  30 | A figure taken (which represents the internal faint nebula), but no description. | 457 |
|  |  | 33.1 |  7  17 | A most wonderful object ; a nebula v B ; v L ; l E ; v g m b M, of an elliptic figure, cut away in the middle by a perfectly definite straight cut 40' broad ; pos = 120°.3 ; dimensions of the nebula, 5' by 4'. The internal edges have a gleaming light like the moonlight touching the outline in a transparency. | 454 |

**Figure 5** - The entry in the Dunlop (1828) catalogue for NGC 5128 discovered on 29 April 1826. The left column gives the Dunlop number, followed by the Right Ascension (1826), the south polar distance and a description of the object. The south polar distance of 47 45 corresponds to a declination of $-42^{o}$ 15'. The number 7 at the end of the description indicates that the galaxy was observed seven times.

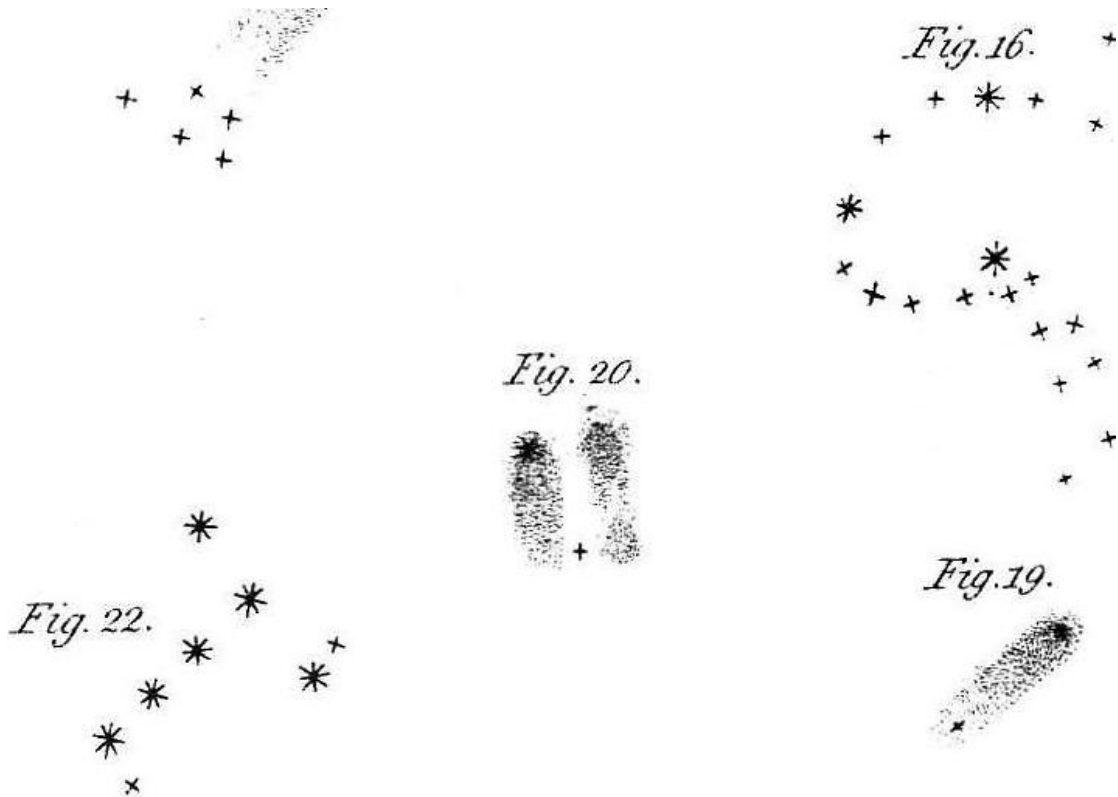

**Figure 6** - Dunlop's (1828) sketch of his nebula number 482 (NGC 5128) is labeled Fig. 20. It is clear from Dunlop's description that he was completely misled by the familiar dark dustband that lies across the object. Dunlop believed that NGC 5128 consisted of two independent nebulae of similar shape that are coincidentally positioned side by side.

Dunlop's position for NGC 5128 is 10.8 arcmin from the centre of the galaxy in position angle 216 degrees. This is mainly because his south polar distance is 8.7 arcmin south of the centre. Later John Herschel found that the positions of objects in the nebulae and cluster catalogue in general were not accurate. The mean distance from the European Southern Observatory Catalogue ESO(B) positions to Dunlop positions is 9.13 arcmin with standard deviation 5.10 arcmin. Dunlop's offsets in Right Ascension in most cases are larger than his offsets in declination and there are no consistent trends.

The speculum mirror in Dunlop's 9-inch telescope had a reflectivity of about 63%. Contemporary aluminium mirrors have a reflectivity of about 87%. This means his 9-inch mirror had the light gathering capacity of a modern 6.5-inch Newtonian and so his limiting stellar magnitude was approximately 13. The telescope's working magnitude limit can be found by analysing and comparing the 60 galaxies Dunlop observed with those galaxies in the LEDA database [see http://leda.univ-lyon1.fr]. Several different methods for analysing the working magnitude limit indicate a limit of 10.9. This was also the magnitude at which Dunlop catalogued 50% and missed 50% of extragalactic galaxies in the LEDA database.

The other instruments used by Dunlop in making these catalogues are not known. However, he may possibly have borrowed a clock from the Parramatta Observatory. Similarly, there is little information regarding the observing techniques used by Dunlop, but he most likely slowly swept by hand back and forth from north to south along the meridian with his 9-inch telescope in sweeps of varying length in south polar distance, a technique he learnt earlier at the Parramatta Observatory. His field of view was approximately 45 arcmin. As he swept he noted the time, probably using a sidereal clock, when a nebula or star cluster crossed the meridian. He also wrote a description and noted the south polar distance using a vernier scale attached to the telescope. Reading the instruments with a candle and writing the description would no doubt have compromised his dark adaptation. Dunlop also had to climb up and down a ladder to observe objects above altitude 40 degrees.

### 2.3  The Naming of NGC 5128

In 1827 Dunlop returned to Scotland where he reduced and wrote up his observations of the southern nebulae and star clusters, probably between August and November. This task was rather poorly executed as a comparison of his written notes (available on microfilm at the National Library of Australia in Canberra) with the final printed catalogue reveals numerous copying errors; for example 13 was copied instead of 30 and 15 instead of 50, possibly dictation errors caused by the person typing the catalogue having difficulty in understanding Dunlop's Scottish accent.

The catalogue of 629 nebulae and star clusters was presented to the Astronomical Society of London on 20 December 1827 and later published in the *Philosophical Transactions* (Dunlop 1828). The second Dunlop (1829) catalogue consisting of 253 double stars was presented to the Astronomical Society six months later in May 1828.

In recognition of this work, the President of the Astronomical Society, Sir John Herschel, presented Gold Medals to Sir Thomas Brisbane and James Dunlop in February 1828. Herschel praised Dunlop for his zealous, active, industrious and methodical work, and also said Dunlop "… must be regarded as the associate rather than the assistant of his employer; and their difference of situation becomes merged in their unity of sentiment and object" (cited in Service 1890). Herschel also noted that "… the astronomers of Europe may view with something approaching to envy, the lot of these their more fortunate brethren". It appears James Dunlop was not present for the presentation of his Gold Medal, possibly due to his low social class, as Herschel asked James South to "… transmit to him [Dunlop] also this our medal". While Herschel and Brisbane were both upper-class gentlemen, Dunlop certainly was not.

In November 1833 Sir John Herschel sailed from England to Cape Town in South Africa where, over the next four years, he would continue the pioneering observations made by Dunlop in Australia. Herschel also followed in the footsteps of the Frenchman Nicolas Lacaille who made the first catalogue of southern clusters and nebulae in 1751–52, also from Cape Town. Lacaille used a miniature 0.5-inch (1.3 cm) refractor to observe 42 objects, most of which were clusters, although there was one galaxy (M83).

Herschel surveyed the southern skies with an 18.5 inch (47 cm) aperture telescope, initially using Dunlop's catalogue of 629 objects as a reference. In early June 1834 Herschel made four observations of Dunlop's nebula number 482 and recorded its coordinates using north polar distance which equated to a declination of –42° 8' 3" (see Figure 7). Despite his success with this nebula, Herschel was unable to find as many as two-thirds of Dunlop's nebulae and star clusters and he soon cast the catalogue aside, criticising it as inaccurate.

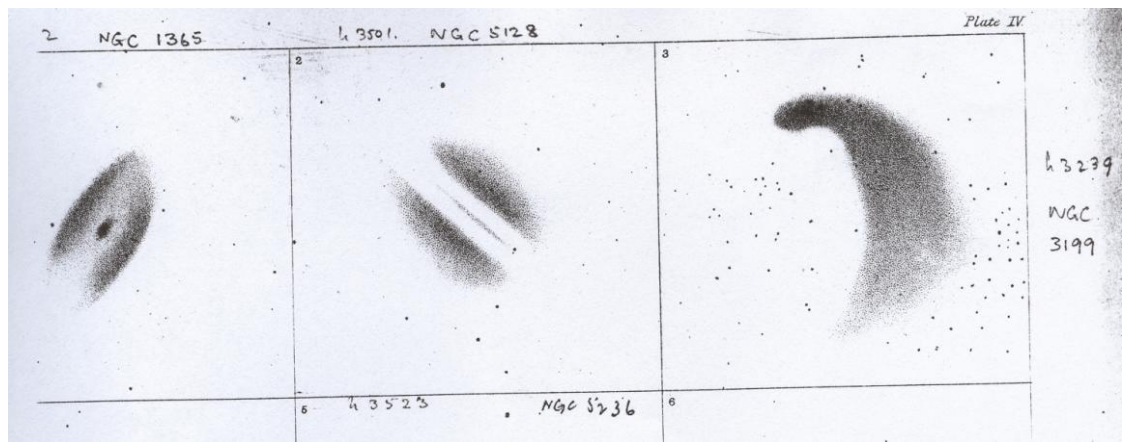

**Figure 7** - John Herschel's drawing of NGC 5128 made in June 1834, showing a faint line in the dark lane [after Herschel 1848].

Herschel (1848) eventually produced his own catalogue of southern nebulae and clusters, in which he noted that his object number 3501 was the same as Dunlop 482. When he later compiled the General Catalogue, the object was recorded as GC 3525. In 1888 John Louis Emil Dreyer expanded the General Catalogue to produce the New General Catalogue, allocating NGC 5128 to this galaxy. In the New General Catalogue, Dreyer only included objects from the Lacaille and Dunlop catalogues if they also had been observed by Herschel. Thus, in 1888, sixty-two years after it was discovered by Dunlop, NGC 5128 finally gained its present optical name.

## 3 Discovery of Centaurus A at Dover Heights

### 3.1 The Radiophysics Laboratory

The Radiophysics Laboratory was established in 1939 under the supervision of the Council for Scientific and Industrial Research (CSIR) to design and construct radar equipment for use by the Armed Services. The initial focus was on the adaptation of the British designs of Shore Defence radars (ShD) and the overseas version of the Chain Home Low (COL) systems for Australian operations. These units were intended to provide early warning of the approach of enemy aircraft and naval ships. Some of the first installations were erected to provide protection for the cities and more vulnerable targets. Two such radars were erected on Collaroy Plateau (COL) and the 75-m high cliffs at Dover Heights (ShD) to scan the air and sea approaches to Sydney. Figure 8 is a general view of the Dover Heights site in 1943, looking north to the entrance to Sydney Harbour with the 200 MHz ShD antenna in the foreground. The site consists of a succession of low sand hills, partially covered in native vegetation.

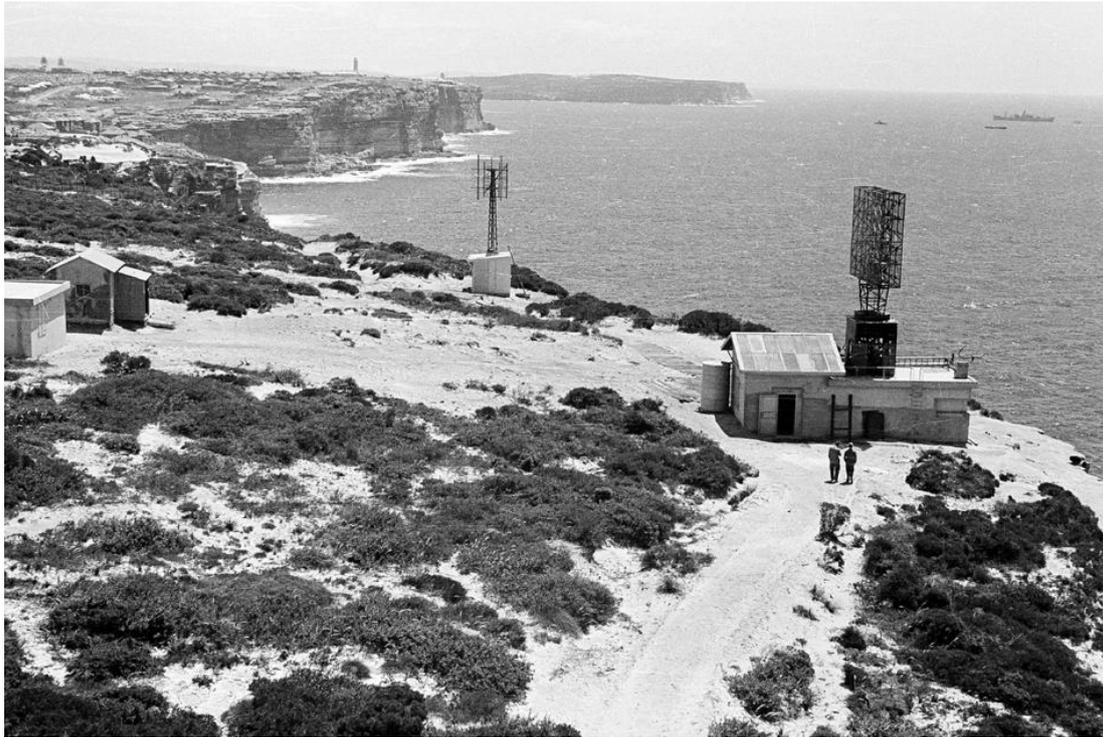

**Figure 8** - The Dover Heights site in February 1943, south of the entrance to Sydney Harbour.

Near the end of WWII in 1945, the staff at Radiophysics was encouraged to use its considerable expertise in physics and radio engineering to propose projects unconnected with warfare. Some of the more astronomically minded staff including Joe Pawsey, Lindsay McCready and Ruby Payne-Scott were already aware that British, European and New Zealand early warning coastal radars had occasionally detected rather intense increases in the 'grass' on their radar displays, often correlated with the rising or setting Sun that happened to be passing through their sea interference fringes (see e.g. Orchiston 2005). An exhaustive investigation of the British reports was published by J. S. Hey (1946), who established with certainty that intense metre-wave solar emission was responsible for these occurrences.

*3.2 Early Observations at Dover Heights*

In late 1945 and early 1946, McCready, Pawsey, & Payne-Scott (1947), using the ShD radar array at Dover Heights and a modified radar receiver and chart recorder, were able to position the solar sources within the sea interference fringes to show that compact sources of intense emission had positions coincident with large sunspot groups in the solar photosphere. It was at these same active intervals that Slee, working with a COL radar at Darwin, independently observed strong solar fringes from the setting Sun (Orchiston & Slee 2002a).

Hey, Parsons, & Phillips (1946), while surveying the Cygnus region of the sky at 64 MHz, discovered a variable radio source in Cygnus, but their angular resolution was too poor to position the source with enough accuracy to identify an optical counterpart. Hey et al. (1946) made no attempt to study the temporal behaviour in any detail, but its rather small angular size of less than one or two degrees prompted

John Bolton to consider the possibility of it being a point-like source (personal communication Bolton to Slee).

Bolton, who had joined Pawsey's then small group in September 1946, made an attempt, while observing the Sun at Dover Heights in November 1946 with a small pair of Yagi antennas, to find sea interference fringes from the Cygnus region, but found that his receiver sensitivity and stability were insufficient for the task. He returned to the Radiophysics Lab in early 1947, where he joined Gordon Stanley in constructing better receivers and higher-gain Yagi antennas. Figure 9 shows three of the main players in establishing the Dover Heights site.

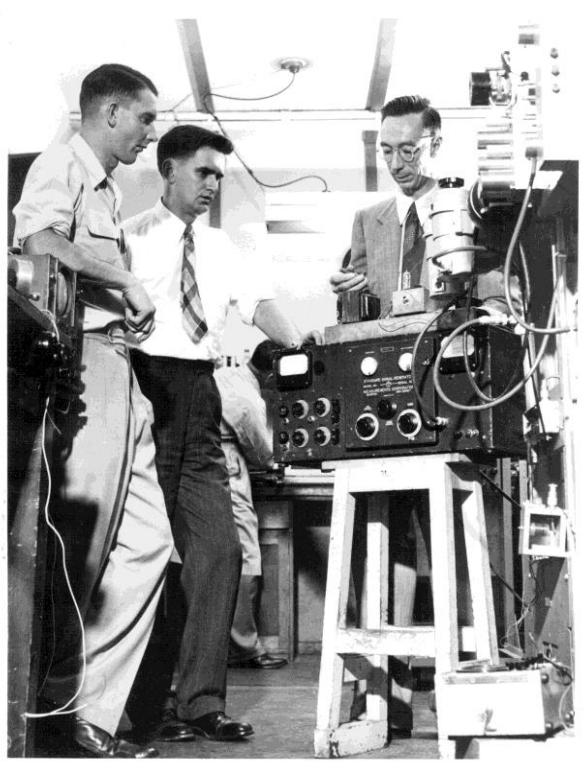

**Figure 9** - Three of the principal figures in early Australian radio astronomy (left to right): John Bolton, Gordon Stanley and Joe Pawsey.

Bolton and Stanley returned to Dover Heights in March 1947, by which time the ShD array had been removed, and mounted pairs of Yagis on the blockhouse to operate at 60, 100 and 200 MHz. For the first month or so they tracked the Sun on a daily basis and were lucky enough to catch a major solar outburst in March 1947. This proved to be the first recognized Type II solar burst, showing the typical slow frequency drift from high to low frequencies (see Payne-Scott, Yabsley, & Bolton 1947).

Sea interference fringes were a regular feature of coastal radar displays. They made their presence known by the regular fading of echoes as the target aircraft moved through the fringes as it approached the radar station. The explanation for this is outlined in Figure 10, which shows that the radio waves reflected by the target reach the radar antenna via two paths: a direct ray from the target and an almost equally strong sea-reflected wave. These two rays interfere with each other, producing maximum intensities when their path-lengths differ by an even number of half-wavelengths and minimum intensities when the path-lengths differ by an odd number of half-wavelengths.

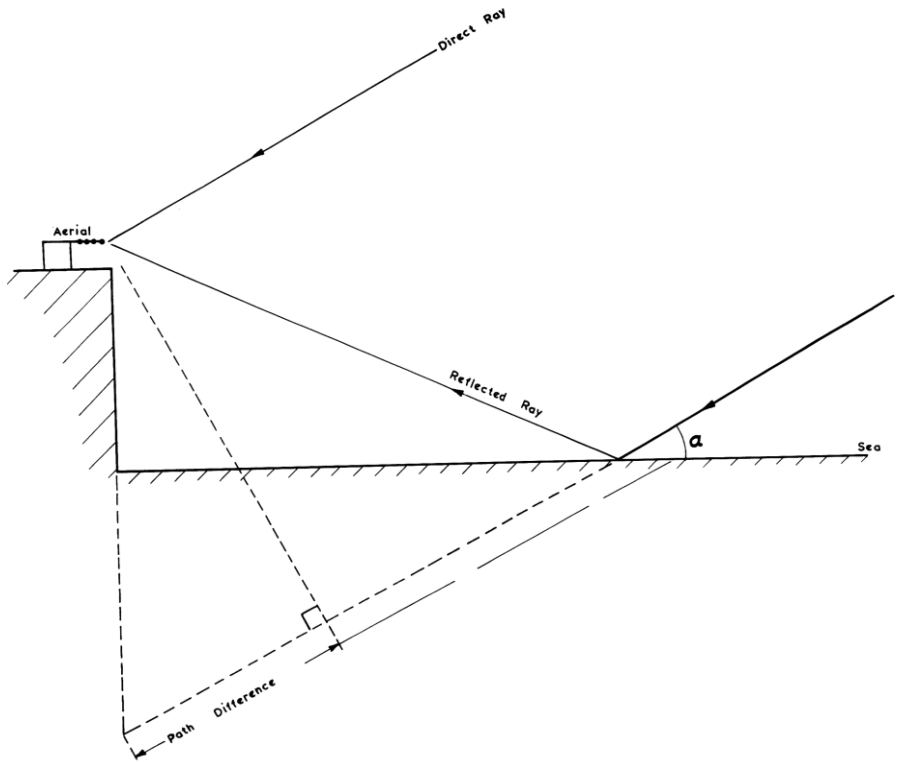

**Figure 10** - Ray tracing in the theory of the sea interferometer. A virtual image of the cliff-top aerial is formed at a point the same distance below sea level.

When applied to the detection of a discrete radio source, the cliff height, operating wavelength and angular size of the source are important parameters. The height of the antenna (expressed in wavelengths) above sea level determines the angular spacing between fringe maxima and minima: the greater the height the better the angular resolution. If, however, the angular diameter of the source begins to approach the fringe spacing, incomplete interference between the direct and reflected waves results in reducing the fringe maxima and filling in the fringe minima. The ratio of the observed fringe amplitude to that of a point-like source provides a measure of the angular diameter of the source.

The first observations of the Cygnus region were made in late May 1947, using the 100 MHz antenna depicted in Figure 11 with the Yagi elements parallel to the horizon and 75 m above the Tasman Sea directly below. The fringes detected on the first night were confirmed on subsequent nights in May and June.

Panel A in Figure 12 shows a typical Cygnus A interference pattern at the time of the discovery, with the intensity variations on the maxima first noted by Hey et al. (1946). For comparison, trace B is a pattern from the quiet Sun using the same equipment and sensitivity; here we see the influence of the ~0.5° diameter of the sun filling in the fringe minima. It is clear that the angular diameter of Cygnus A is much less than that of the Sun. The details of this important discovery were first published in *Nature* by Bolton & Stanley (1948).

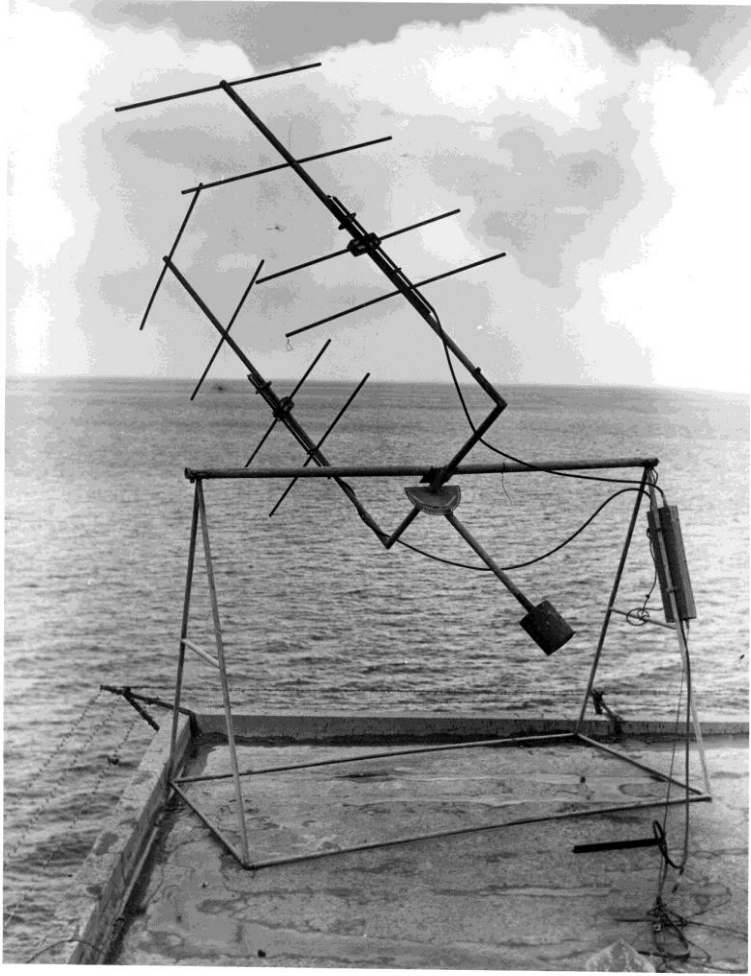

**Figure 11** - The pair of 100 MHz Yagis used to find the first fringes from Cygnus A. For sea interferometry the Yagi elements are turned parallel to the horizon.

### 3.3 The Detection of Centaurus A, Virgo A and Taurus A

During late 1947 and early 1948, Bolton, Stanley and Slee conducted a general survey of the sky with the same antenna and receiver used to obtain the fringes in Figure 12. The successful observations of Cygnus A had suggested that it should be possible to detect sources with flux densities as low as one-tenth that of Cygnus A, provided that similar sources existed and that their angular diameters were much less than a degree. They were helped in this survey by the use of much better AC power stabilisation. A survey of the horizon from declination $-60°$ to the northern limit of $+50°$ with an antenna beamwidth of $25°$ in $5°$ steps would take a minimum of 22 days, even if the weather was good and there were no equipment problems. Further delay was introduced by the restriction to nocturnal observation to reduce city and solar interference, and to cover the full range of right ascension.

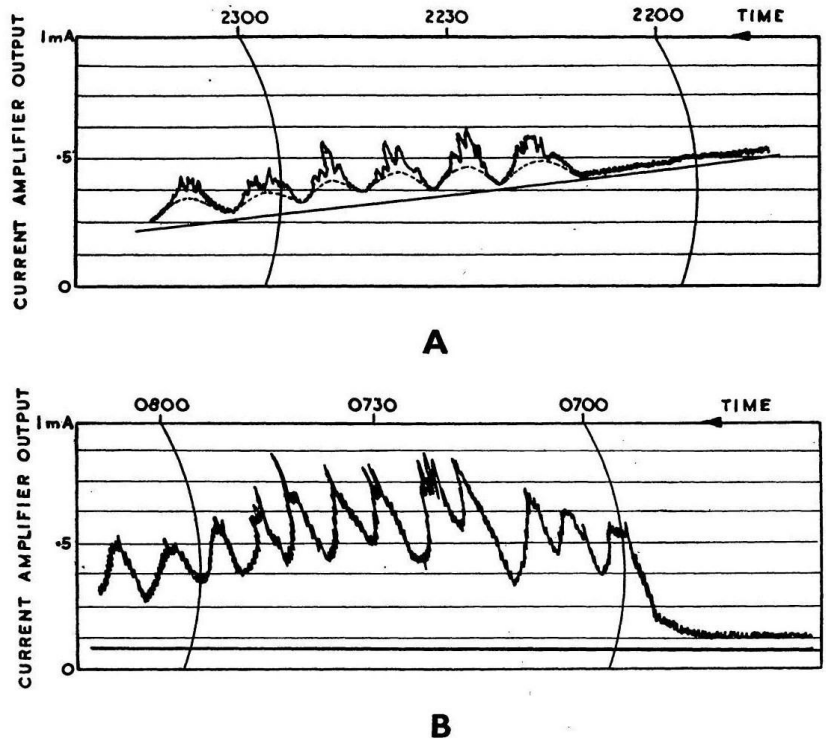

**Figure 12** - Panel A: the discovery fringes of Cygnus A rising after 10 pm in the evening (from right to left). Panel B: the fringes from the quiet Sun detected using the same equipment.

If signs of an interference pattern were detected, the observations at that particular rising azimuth would be repeated, sometimes for two or three successive nights. These observations were trying for the duty observer as the total power deflection on the recorder chart varied according to the galactic latitude, so that frequent adjustment of the recorder pen was often necessary. (It was not until 1952 that this problem was solved as explained in Section 5.)

During a four month period from late 1947 to early 1948, three new discrete sources, which were later named Taurus A, Virgo A and Centaurus A, were unambiguously detected. The discovery fringes for Taurus A are shown below in Figure 15, which is the only archived recording of this trio of important sources. Centaurus A and Virgo A would have provided similar fringe amplitudes as their 100 MHz flux densities are similar to Taurus A.

## 4 The New Zealand Expedition

By the end of 1947 the Dover Heights group had recorded Cygnus A, Taurus A, Virgo A and Centaurus A, with evidence of possibly three further discrete sources. In Bolton's (1982) own words "1947 had been a vintage year". However, one overriding problem remained: to establish the true nature of these radio sources far more accurate positional data would be required before they could be identified with objects known to optical astronomers. With this objective in mind, Bolton and Stanley began searching for a better observing site, one that would permit observations of both the rising and setting of radio sources. They investigated elevated coastal locations close to Sydney, sites on both Lord Howe and Norfolk Islands to the north-east of Sydney, before deciding on the peninsula region near Auckland on the north island of New Zealand (see Figure 13).

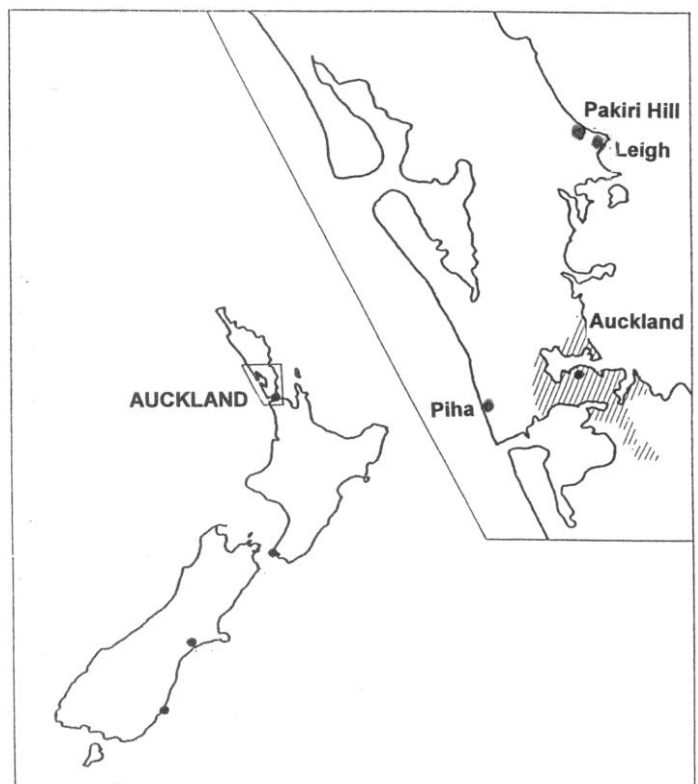

**Figure 13** - Location of the observing sites used by Bolton and Stanley during the New Zealand expedition in June–August 1948 [after Orchiston 1994].

The chief of the Radiophysics Laboratory, E. G. ('Taffy') Bowen was an enthusiastic supporter of the expedition and early in 1948 he began contacting New Zealand authorities such as the Surveyor-General's Department, the PMG Department and the Auckland Weather Bureau for more detailed information on the suitability of conditions in the area. As Bowen (1948) stated in a letter to the CSIR Executive in March 1948, the purpose of the expedition was:

> "To determine as precisely as possible the positions and upper limits of angular width of the variable sources of cosmic noise already approximately determined by observations from Dover Heights. At present some seven such sources have been found. The accuracy with which these determinations may be made is considerably improved if both the rising and setting of the sources can be observed over a sea path."

At the end of May 1948 a converted ex-Army radar trailer, containing four 100-MHz Yagi antennas, a new receiver, recorders, chronometers and weather-recording equipment, was shipped to Auckland (see Figure 14). Bolton and Stanley followed a week later by air, while Slee continued the observations at Dover Heights in their absence. With the cooperation of the Department of Scientific and Industrial Research (the New Zealand counterpart of the Australian CSIR), an army truck was provided to haul the mobile sea interferometer to its first observing site Pakiri Hill, a farm owned by the Greenwood family, the original settlers in the area, 10 km from Leigh and about 70 km due north of Auckland. The DSIR supplied a ration of petrol, a book of requisitions with which all bills were paid, and even provided a liaison officer to assist during the first two weeks. Further cooperation came from staff from the Department of Lands and Survey who accurately determined the height of the Pakiri site (Orchiston 1994).

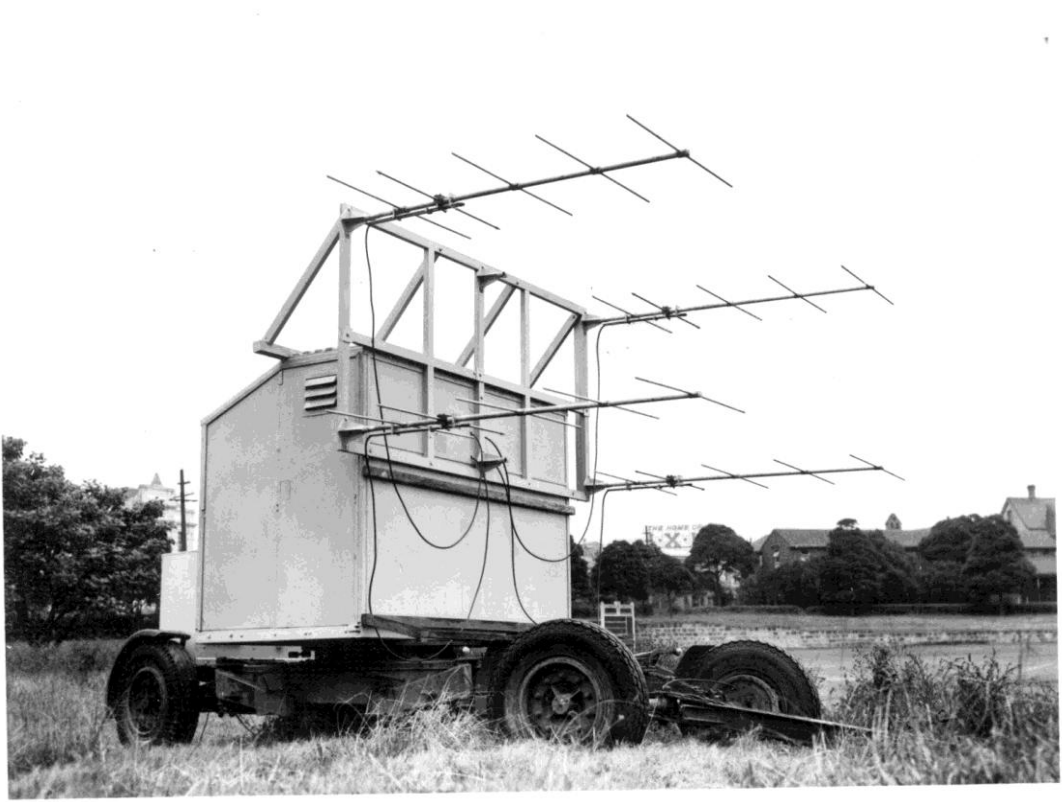

**Figure 14** - Mobile sea interferometer in the grounds of the University of Sydney, shortly before it was shipped to New Zealand in May 1948.

After several days at the site Bolton (1948a) could report to Bowen:

> "Cooperation both official and unofficial has been magnificent. The farmer on whose land we are sited has raised no objection to us using his timber, digging holes in his paddocks etc. – in fact has done everything to assist. They even brought us tea and sandwiches at five o'clock in the morning on the last two nights – for which we are very grateful. Nine hours at a stretch without Dover's comforts is just a little tough."

The site was at an altitude of 300 m, over three times higher than the cliff-top at Dover Heights, thus providing an effective interferometer baseline over three times longer. The coastline ran from south-east to north-west, so that sources could be followed from the time they rose above the horizon. The Yagi array had a horizontal beamwidth of 12° and a vertical beamwidth of 30°.

Bolton and Stanley spent almost two months at Pakiri Hill and endured the physical hardships of camping out in the open during the coldest winter months. As Bolton (1982) later recalled:

> "Conditions were far from ideal; we had a long extension from an already overloaded power line and frequency variations caused variations in the chart recorder speed of at least 10%. The weather was sometimes appalling; on one occasion our barometer recorded a fall of 15 mm in 30 sec, to be followed by a similar fall of 9 mm, 10 min later."

Despite these difficulties, by mid-July about thirty useable observations of Cygnus A and five of Taurus A were obtained (see Figure 15).

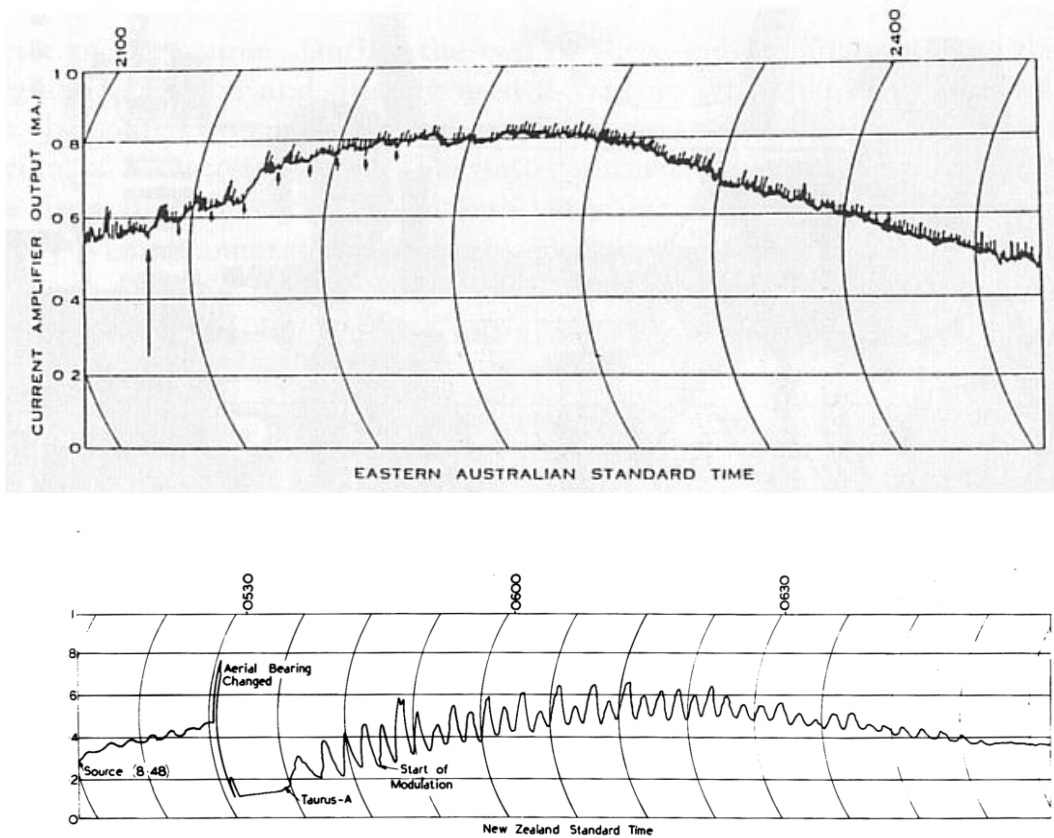

**Figure 15** - The fringes from Taurus A. *Top*: The discovery observation made on 6 November 1947 at Dover Heights. The rising time is shown by the arrow on the left and there are numerous spikes of interference. *Bottom*: The fringes recorded at Leigh in July 1948 illustrate the improved signal-to-nose ratio of the New Zealand observations. The recording also shows the beating pattern between the two sources present – source 8.48 and Taurus A – within the four-Yagi primary beam.

Early in August the trailer was towed to the cliff-top at Piha, the site of a former World War II radar station 30 km due west of Auckland. The site, also at an elevation of 300 m, offered an excellent view of the western horizon where sources could be viewed setting and the data compared for the same sources rising in the east at Pakiri. This meant some of the errors could be minimised, the most important being the amount of atmospheric refraction that the signal from a source undergoes as it rises above and sets below the horizon at different angles. Bolton and Stanley soon discovered that conditions at Piha were much superior to those encountered at Pakiri: "The diesel plant for the radar station provided a supply of electricity stable in both voltage and frequency, our receivers performed faultlessly and the weather was perfect" (Bolton 1982). Over the next two weeks further observations were made of Cygnus A, Taurus A, Centaurus A and Virgo A.

The New Zealand expedition produced three outcomes of major significance to the fledgling science of radio astronomy. First, it was now possible to fix the position of seven sources to an accuracy of about 10 arcmin, a considerable improvement on the accuracy of about 1 degree at Dover Heights reported in the *Nature* paper that appeared in July (Bolton 1948b). In addition, the number of confirmed sources had now risen from 7 to 13. Second, as well as the confirmed sources, there was now compelling evidence for up to fifty new sources. However, because of their relatively low intensity, there had not been enough time during the New Zealand visit to investigate these fainter sources in more detail.

The third notable result from the expedition involved the enigmatic source fluctuations or 'twinkling' in Cygnus A, first reported by Hey, Parsons & Phillips (1946). The New Zealand observations on Cygnus A could be compared those made at exactly the same time by Slee at Dover Heights. These simultaneous trans-Tasman observations provided the first evidence that the puzzling amplitude fluctuations were due to terrestrial atmospheric scintillations, and were not intrinsic to the Cygnus source itself as had been widely believed previously (see Stanley & Slee 1950). In turn, this removed the apparently compelling evidence that Cygnus must be a relatively close galactic object. All in all, the New Zealand expedition was judged to be a great success (Orchiston 1994).

Bolton and Stanley returned to Sydney in mid-August 1948 and began the laborious task of reducing two months of data. Bolton produced a series of brief unpublished reports on each source, providing details on the calculation of both Right Ascension and Declination, the equipment used and the sites where the observations were made. Most of the reports were produced in September but the fifth in the series - devoted to Centaurus A - was delayed until December (see Figure 16). This was because of the difficulty in resolving Centaurus from another nearby source, the same reason why a position had not been given for the source in the paper announcing its discovery (Bolton 1948). In the meantime a new aerial system had been installed at Dover Heights with a much narrower beamwidth than the one used for the discovery observation, and the new aerial was able to successfully resolve the two sources. The position for Centaurus A was calculated by combining observations with the new aerial made in early November 1948 and those made at Piha in early August.

In the final paragraph of the Centaurus A report, under the heading 'Identification of Source', Bolton (1948c) stated:

> "The limits in position of the source RA 13 hrs 22 mins 20 secs $\pm$ 1 min, Declination –42° 37' $\pm$ 8' enclosed NGC 5128. This object is classed as an extragalactic nebula. It is a seventh magnitude object with a peculiar spectrum. [Walter] Baade calls it a freak and it is referred to by [Harlow] Shapley as a 'pathological specimen' though no details are known at present as to the exact nature of its peculiarity. It will be an interesting object to study with the Stromlo nebular spectrograph during the late summer months."

This paragraph not only provided the first measured position of the radio source but, as is now known, correctly identified Centaurus A with the optical galaxy NGC 5128.

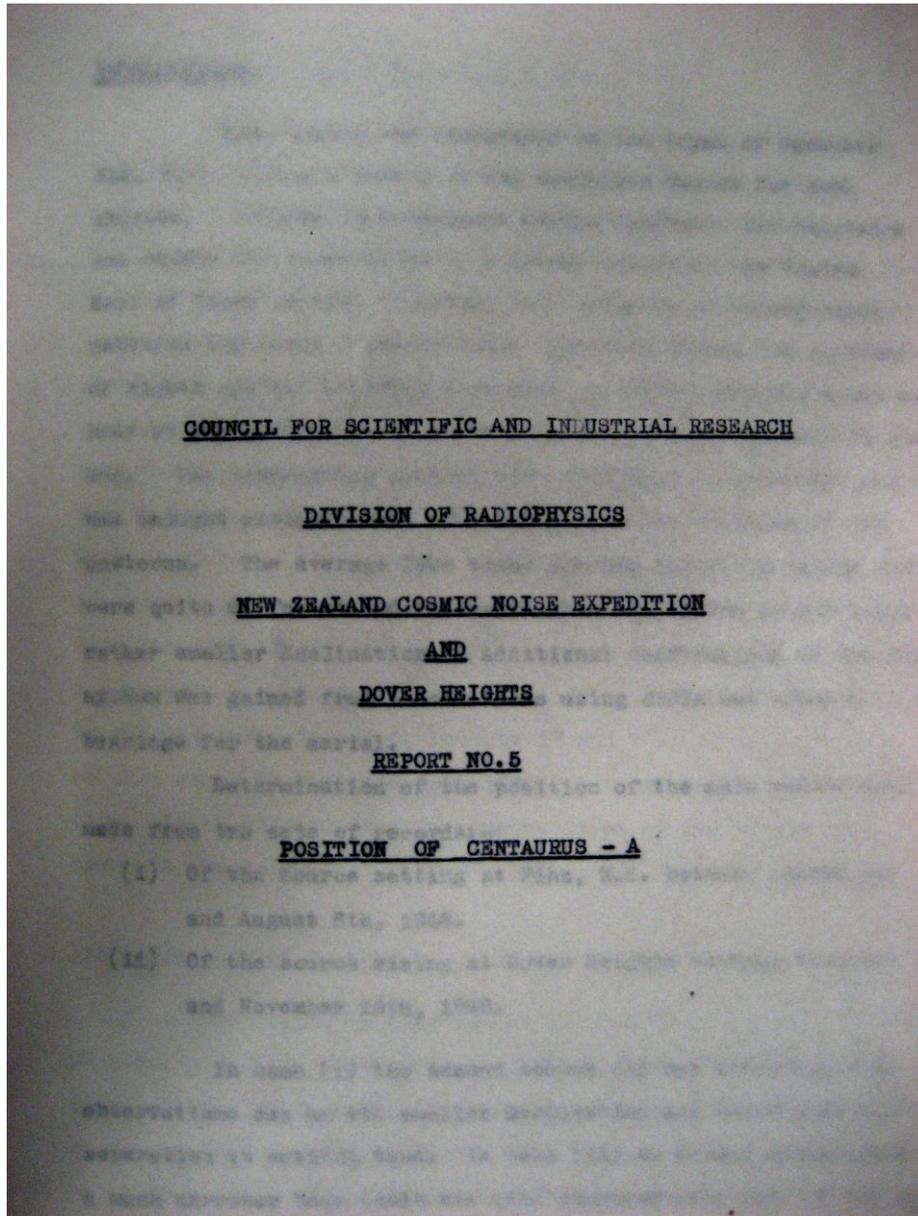

**Figure 16** - An unpublished report by John Bolton in December 1948 provided the first position measurement for Centaurus A and also the evidence for the identification of the source with the optical galaxy NGC 5128 [courtesy National Archives of Australia]

## 5 Further Studies of Centaurus A at Dover Heights

### 5.1 Spectral Measurements

During 1948–49 the first intensity spectra of the four strongest sources were measured using small Yagi arrays at up to five frequencies between 40 and 160 MHz. Stanley & Slee (1950) reported these spectra, which show the variation with frequency of the radio emission from discrete sources.

In Figure 17 we see that three of the sources, Cygnus A, Virgo A and Centaurus A, possess the typical steep decline of intensity with frequency that is characteristic of non-thermal synchrotron emission from radio galaxies. Taurus A was identified as the radio remnant of the bright supernova of 1054 AD (now popularly known as the Crab Nebula). Whereas the three radio galaxies have very similar spectra with intensity inversely proportional to frequency, the Crab Nebula (which is also home to a strong pulsar) possesses a much flatter spectrum.

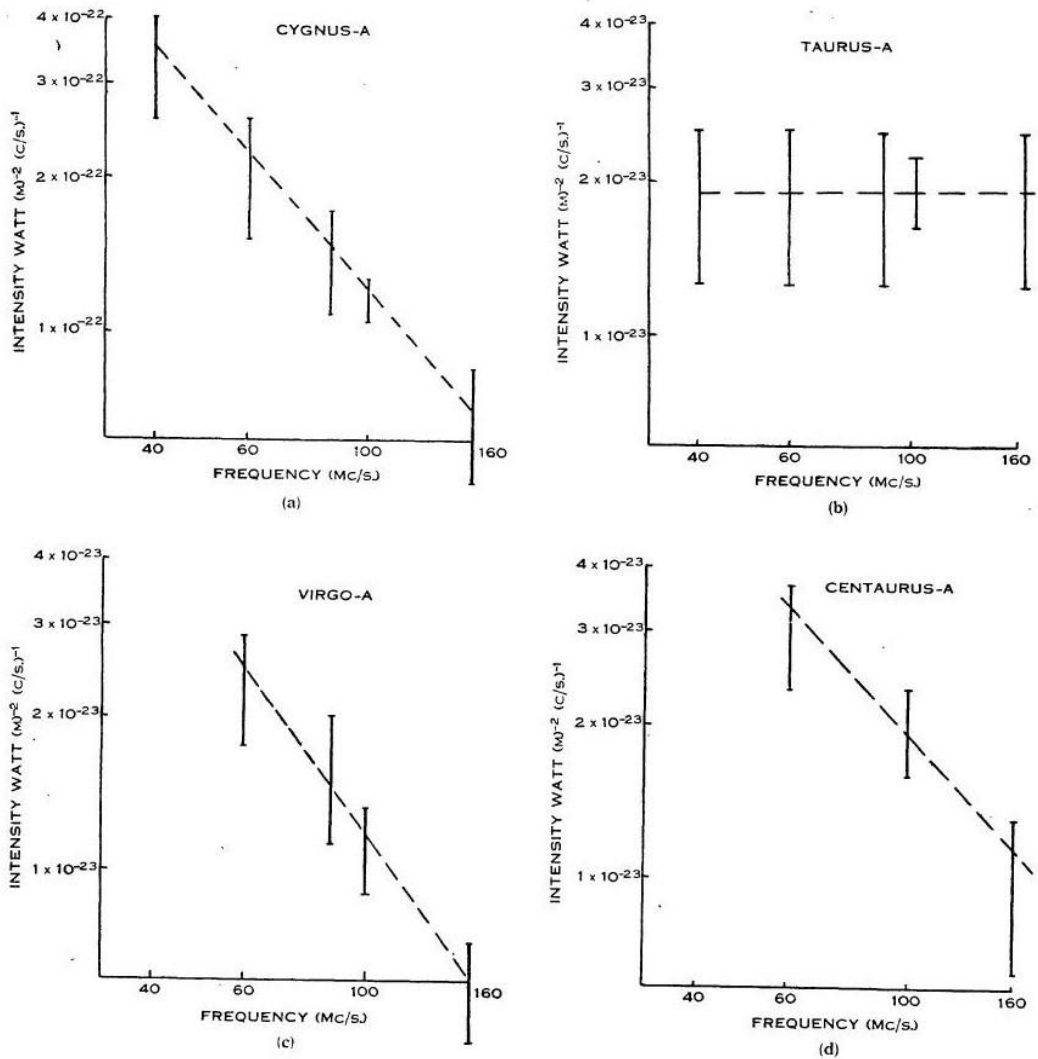

**Figure 17** - The first radio spectra of the four strongest sources discovered at Dover Heights. The intensity measurements were at up to five frequencies with two-Yagi sea interferometers.

The spectral index (given by the slope of the dashed lines) yields basic information about the emission mechanism. The high and almost equal spectral indices of the three radio galaxies are now thought to indicate that these radio sources were formed approximately 100 million years ago. In that elapsed time the highest energy electrons, originally responsible for the emission at the high frequency end of the spectrum, will have lost much of their original energy in the form of synchrotron radio emission. This results in a gradual steepening of the radio spectrum. On the other hand, the Crab Nebula is known to be a very young object with an initially flat synchrotron spectrum, which has not yet had enough time to show the same degree of steepening.

## 5.2 Angular Size Measurements

As well as the intensity spectrum of a source, its angular size (leading to its linear size if its distance is known) is of equal importance in understanding the forces acting on the ionised gaseous structure of the source. For example, successive measurements of the angular diameter of a supernova remnant such as Taurus A, initiated by the collapse of a massive, evolved star, gives a measure of its rate of expansion into the surrounding interstellar medium.

In the case of a radio galaxy such as Centaurus A at a known distance, we now know that its progenitor was a massive black hole, from which jets of ionised gas were ejected to feed the two radio lobes that form the compact central part of the source. A knowledge of its angular size and hence its linear size leads to an estimate of the average velocity of the relativistic electron streams that eventually form the lobes. The existence of the much more extended radio source that surrounds the inner radio lobes was not known until 1951–52, when it was first detected at Dover Heights in a specially designed experiment, first fully described by Bolton et al. (1954). This was followed in 1953 by mapping Centaurus A at 400 MHz (McGee, Slee, & Stanley 1955).

*The compound altitude–azimuth interferometer at Dover Heights.* By 1951 the Dover Heights group, consisting of Bolton, Stanley, Westfold and Slee, decided to measure the angular sizes of the stronger discrete sources. To achieve this aim, a compound interferometer consisting of two four-Yagi 100 MHz arrays was deployed along the cliff edge in a nearly north–south orientation. The observations of Centaurus A were made by pointing the pair of antennas towards the rising azimuth of the source and recording the resulting complex fringe patterns at six spacings of the antenna-pair of up to 21 wavelengths. The outputs of the two small arrays were multiplied by phase switching at 25 Hz to form a Michelson interferometer, while at the same time a sea-interferometer was formed by their vertical spacing above the ocean. The resulting set of six chart recordings is depicted in Figure 18.

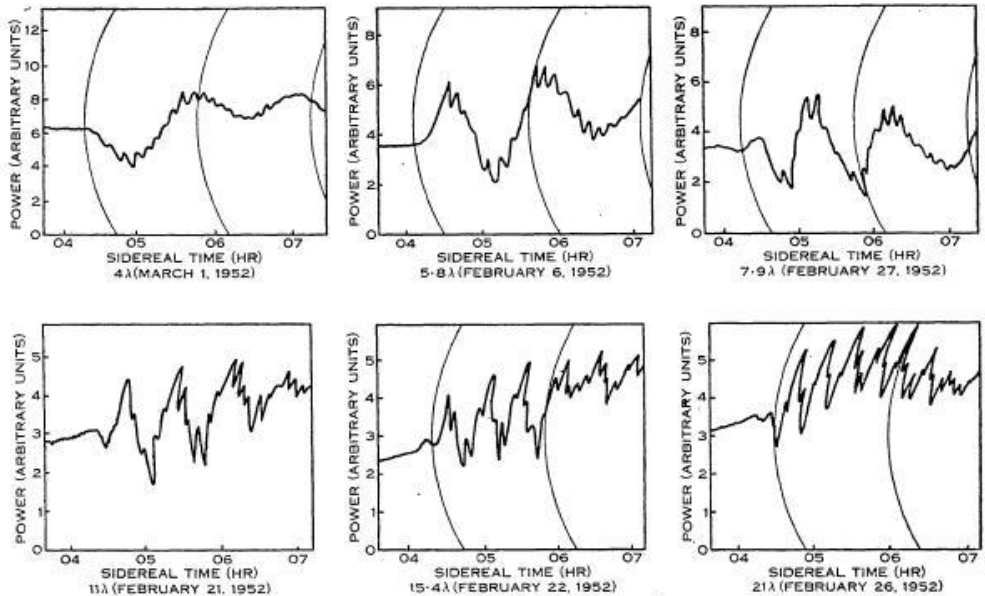

**Figure 18** - Complex fringe patterns formed by phase-switching the outputs of a pair of antennas spaced from 4 wavelengths (top left) to 21 wavelengths (bottom right) along the cliff top - see text for explanation.

The complex fringe patterns in Figure 18 can be regarded as the product of two interference patterns: the first from the short-period sea-interference fringes with constant angular resolution of ~1.1° due to their fixed height of 75 m above sea level, is due to the movement of the source in elevation. The second pattern, from the longer period fringes of the Michelson interferometer, is due to the source's movement in azimuth. The very slow fringes in the top left panel in Figure 18 show the presence of a large diameter source of dimensions considerably less than the Michelson fringe spacing of 14.3°. The amplitude of the slower fringes begins to drop rapidly in panel three where the Michelson resolution is 7.2°, as shown in Figure 19. With the spacing reduced to 15.4 wavelengths (resolution 3.7°) the slower Michelson pattern has all but disappeared, and in the last panel, with an angular resolution of 2.7°, the fringe pattern is almost entirely due to the sea interference fringes. At no stage do these fast elevation fringes disappear, occurring increasingly as a minor ripple on the slower azimuthal fringes as their angular resolution decreases.

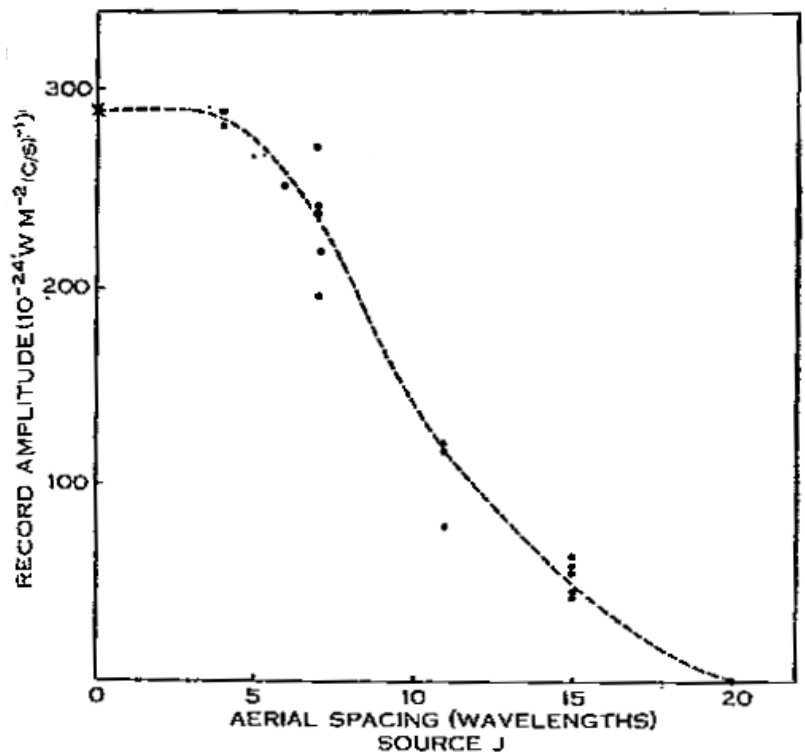

**Figure 19** - Amplitude of the Michelson fringes (azimuthal movement) for Centaurus A as a function of antenna spacing in wavelengths.

This experiment demonstrated for the first time that the more compact inner sources of Centaurus A are surrounded by an extended object of some 4–5° in the azimuthal direction, which contributes most of the flux density at metre-wavelengths. The total flux density of Centaurus A at 100 MHz is ~3 × 10$^4$ Jy and equal to that of Cygnus A, which hitherto had been accepted as the strongest discrete source in the southern sky, although because of its much larger distance, the absolute radiated power of Cygnus A is many orders of magnitude greater.

*The 21.9-m hole-in-the-sand antenna at Dover Heights.* In late 1951 Bolton, Stanley and Slee spent some hundreds of their lunch-time hours excavating a 21.9-m hole in the sand approximately 150 m to the north of the blockhouse shown in Figure 8, with the aim of building an experimental transit dish to operate at 160 MHz (Orchiston & Slee 2002b). The parabolic shape of the depression was formed by the use of a rotating wooden jig and the reflector, consisting of metal binding strips from large shipping containers, was pegged to the surface with an East–West elongation, which would dictate the accepted polarized component. A guyed metal mast, on which a simple 160 MHz dipole and reflecting element was mounted, could be swung up to 35° in the meridian plane by hauling on the guy ropes. The focal feed was connected to a simple 160 MHz superheterodyne receiver, which used very effective power stabilisation, and the various transit scans were recorded on a chart recorder. A low-noise preamplifier was located in a weather-proof housing at the base of the mast and connected to the rest of the equipment in a nearby mobile trailer.

To test the dish, receiver and concept as a whole, a 160 MHz survey of the Galactic Plane was made with the resulting 6° beam over a 35° range of galactic longitude and ± 10° of galactic latitude surrounding the zenith declination of –34°. Clear contours of a strong 160 MHz source were found at a position close to that of the source now known as Sagittarius A (or Sgr A) in early 1952, as shown in Figure 20.

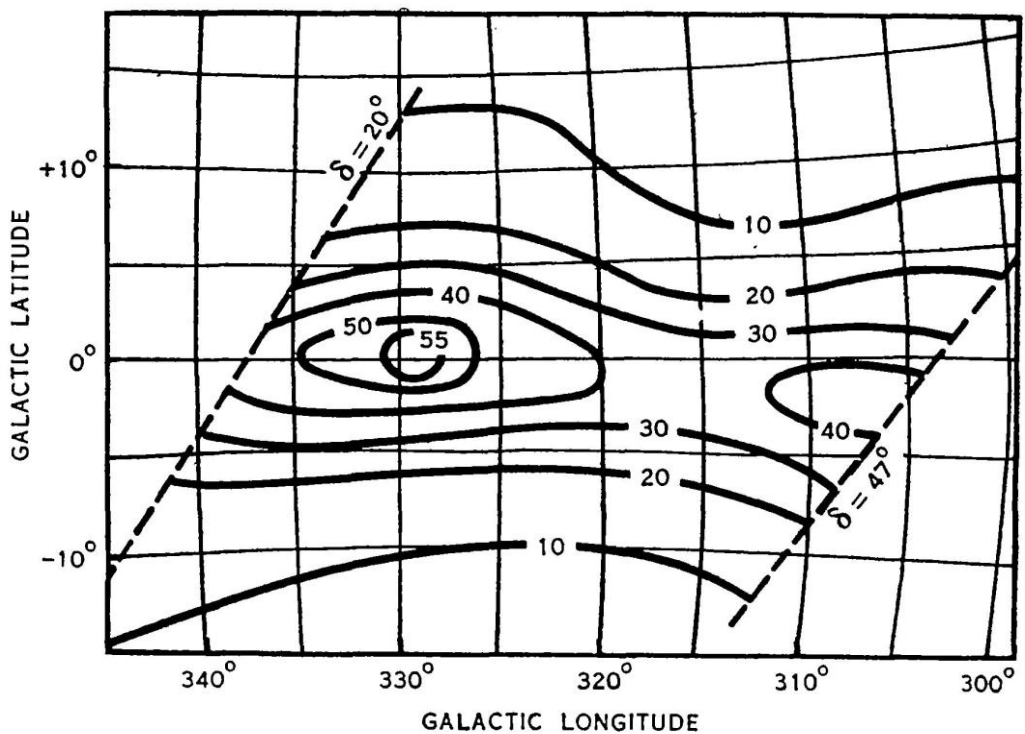

**Figure 20** - Contours of antenna temperature of the region about the Galactic centre as recorded by the 21.9-m hole-in-the-sand dish in early 1952. The units are not accurately known but the contour interval is about 20 K.

Sgr A was first detected at 1210 MHz by Piddington & Minnett (1951). These 160 MHz exploratory contours of Sgr A were made and their source recognised in early 1952, but were not published until later by McGee & Bolton (1954). However, we can say with some certainty that the first metre-wave detection of Sgr A was made in early 1952, using this preliminary 21.9-m dish.

*The final 24.4-m hole-in-the-sand antenna.* Since the proof-of-concept observations in early 1952 had proved so successful, steps were taken in 1952 to construct a slightly larger dish with a more accurate reflecting surface, a broad-band 400 MHz focal feed and a more sophisticated receiver with accurate calibration hardware.

The refurbished dish was constructed using the resources of the well-equipped Radiophysics workshop. This time, an accurately constructed jig was used to shape the parabolic surface, which was concreted and covered with a fine-wire mesh to serve as a high-quality reflector. The sky survey was conducted by swinging the long metal focal mast in the meridian plane by means of high-quality nylon guy ropes to cover the declination range −17° to −49° and making transit scans at 0.5° intervals of zenith angle. Figure 21 shows the completed 400 MHz dish.

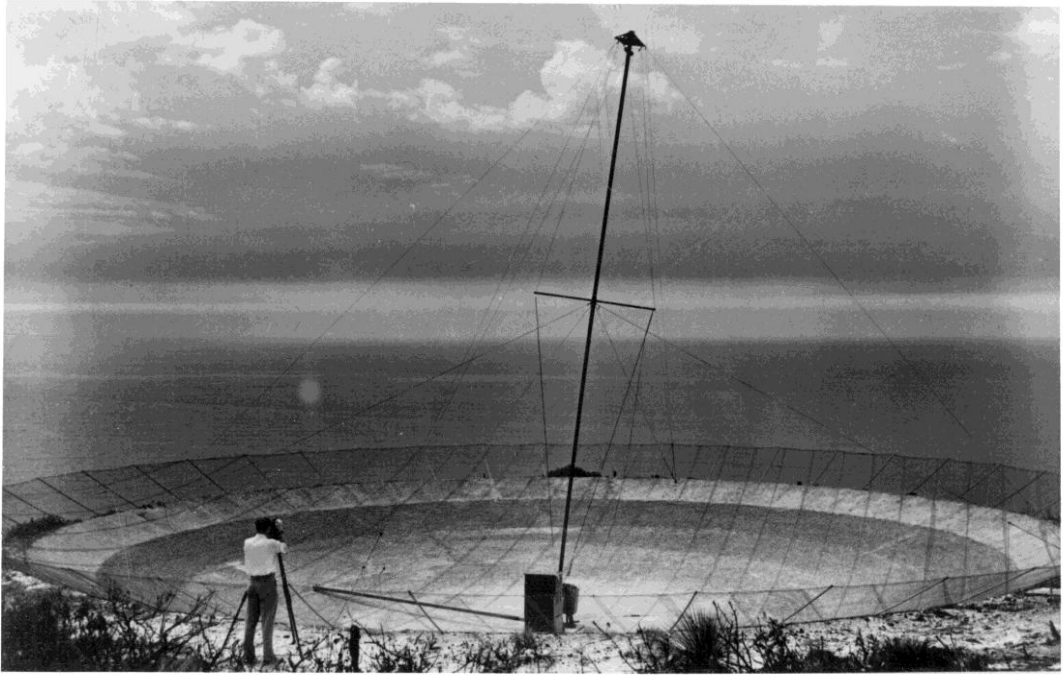

**Figure 21** - The enlarged concrete-surfaced 24.4-m dish during the survey in 1953, with Gordon Stanley using a theodolite to set the zenith angle of the antenna mast.

The equipment and survey results were described in detail by McGee & Bolton (1954) and McGee et al. (1955). Although the main purpose of the improved 400 MHz dish was to survey the Galactic Plane on either side of Sgr A, scans were made of several sources including Centaurus A, which lay well within the declination coverage of the dish. The contours of this source are presented in Figure 22.

It is clear from Figure 22 that the angular diameter of the halo is about 6°, agreeing with its diameter derived by the 100 MHz compound interferometer discussed above. The inner lobes of Centaurus A are not well resolved, but their orientation in position angle of about 45° East of North agrees approximately with later resolved images (see e.g. Slee 1977), which shows the lobes extending some 15 arcmin in the same position angle.

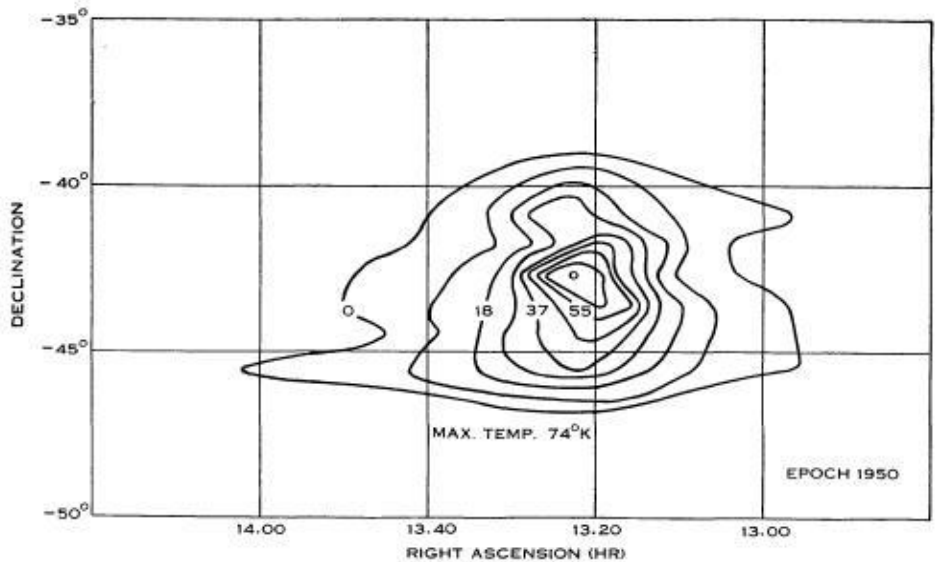

Fig. 5.—Contours of equivalent aerial temperature at 400 Mc/s of the discrete source Centaurus–A. Contour interval is 9·25 °K. (Not corrected for aerial beam.)

**Figure 22** - Contours of equivalent antenna temperature at 400 MHz for Centaurus A using the 24.4-m hole-in-the-sand dish with a 2° beamwidth. The effects of beam smoothing have not been removed.

### 5.3 Other Discrete Source Surveys at Dover Heights

Discrete source surveys were made with increasingly improved antennas and receivers for most of the eight-year life of the field station. The initial survey in late 1947 and early 1948 with a two-Yagi antenna and simple total power receiver discovered the four brightest radio sources south of declination 50° North. In 1949 Stanley & Slee (1950) used a nine-Yagi array mounted on the blockhouse and a much more stable total power receiver to extend the Dover Heights source list to the next tier of sources with the detection of Fornax A, Pictor A, Puppis A and Hercules A, all of which were four to five times weaker with 100 MHz flux densities in the range 200–300 Jy. Interestingly, three of these sources were later identified with powerful radio galaxies and Puppis A with another supernova remnant. Strangely, the Dover Heights group did not unambiguously discover the strong radio galaxy Hydra A, although a number of other sources were listed which were probably causing confusion near the rising azimuth of Hydra A; however, these sources were not confirmed by a later survey.

In 1952–53 the final 12-Yagi array with a primary angular resolution of 8°×12° was erected on the cliff edge and connected to a more sophisticated receiver (see Figure 23). The receiver used with this antenna incorporated an electronic Dicke switch, which improved the stability of the receiving system considerably. During this final survey a method was used for controlling the highly variable background deflection on the chart recorder, permitting unattended operation of the survey (see Bolton & Slee 1953). This led to a remarkable improvement in fringe visibility, stability and reproducibility, as shown in Figure 24. The final survey with the 12-Yagi array produced a list of 104 sources with 100 MHz flux densities as low as 50 Jy. Fifty-one of them were fully or partially confirmed in other surveys done by Bernard Mills with the Badgerys Creek interferometer, as discussed in the next section.

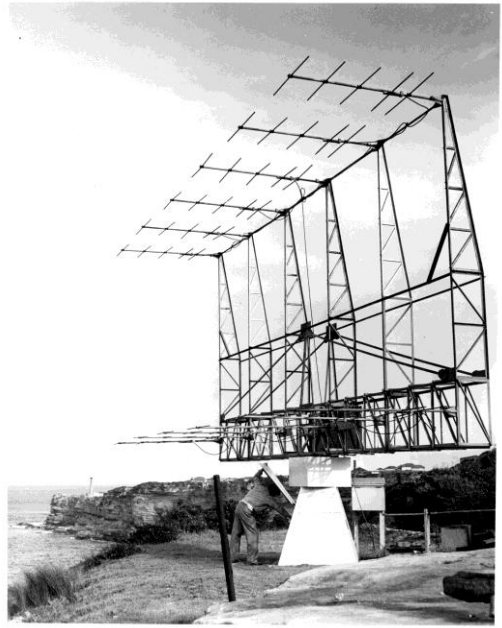

**Figure 23** - The final 12-Yagi 100 MHz antenna at Dover Heights with John Bolton checking the receiving equipment at the base.

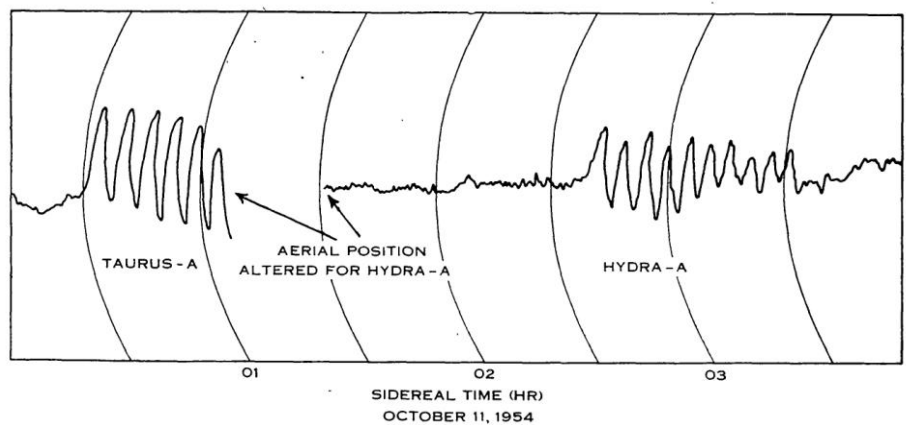

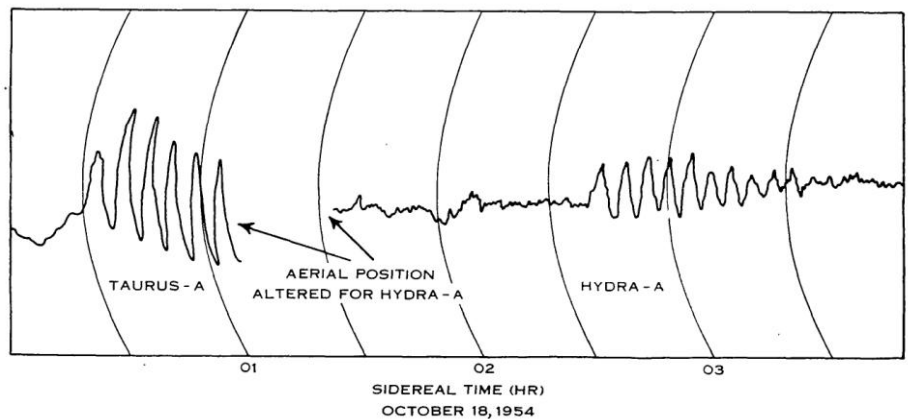

**Figure 24** - Fringes from Taurus A and Hydra A on two dates in October 1954. The improvement in visibility is clear when comparing these fringes of Taurus A with the early sea interferometer recordings – see Figure 15.

## 6  Other Observations by the Radiophysics Lab to 1960

Apart from the ground breaking work at Dover Heights, different teams working at the other CSIRO Radiophysics field stations (see Figure 25) also made significant contributions to elucidating the nature of Centaurus A.  These contributions are discussed in the following section.

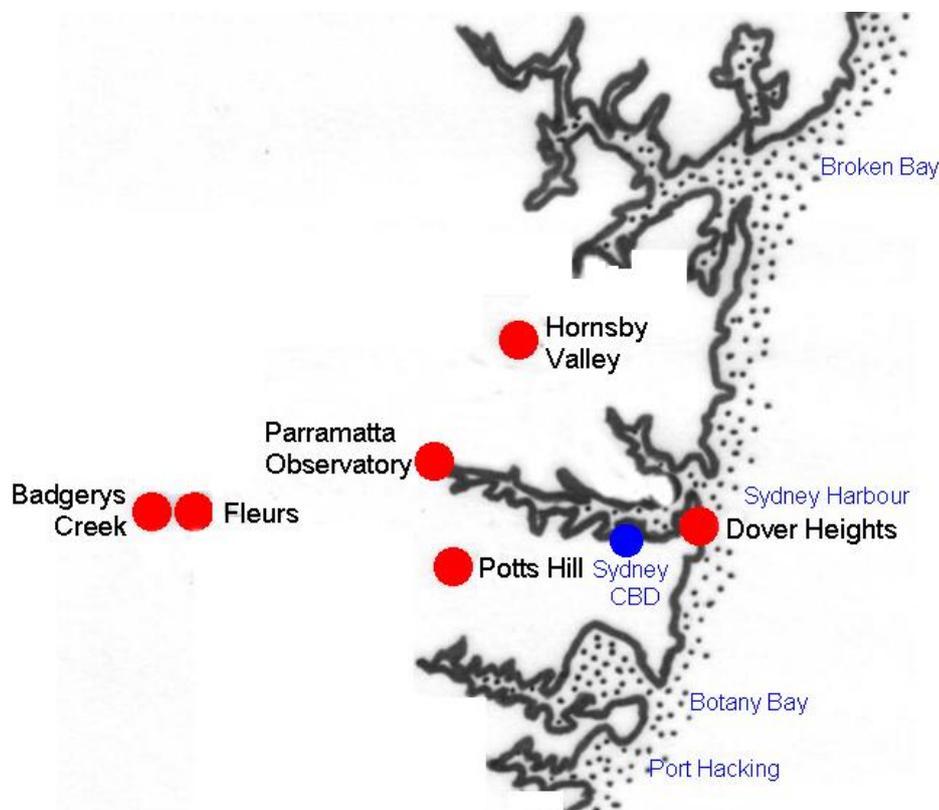

**Figure 25** - The location of the Radiophysics field stations discussed in the text within the greater Sydney region [after Orchiston & Slee 2005].

### 6.1  Potts Hill—1951

The Radiophysics Potts Hill field station was established in 1948 on banks of one of Sydney's major water storage reservoirs. The initial aim of research at the field station was to investigate solar radio noise.  During 1950, Jack Piddington and Harry Minnett (1951) used a 16-ft × 18-ft (6.3 m × 7.1 m) paraboloid (see Figure 26) to investigate discrete sources of cosmic radio noise. This paraboloid had originally been used for solar observations and was also later used by W. N. 'Chris' Christiansen and Jim Hindman to confirm the 21-cm hydrogen-line discovery in July 1951.  As part of their discrete source investigations, Piddington and Minnett were able to extend the observed radio spectrum of Centaurus A to higher frequencies.  Observing at 1210 MHz with a beamwidth of 2.8°, they measured a flux density for Centaurus A of 410 Jy (4.1 × 10⁻²⁴ W m⁻² Hz⁻¹) with an uncertainty of 30%.  During these same observations, Piddington and Minnett also discovered the discrete radio source Sagittarius A (referred to as the Sagittarius–Scorpius source in Figure 27).

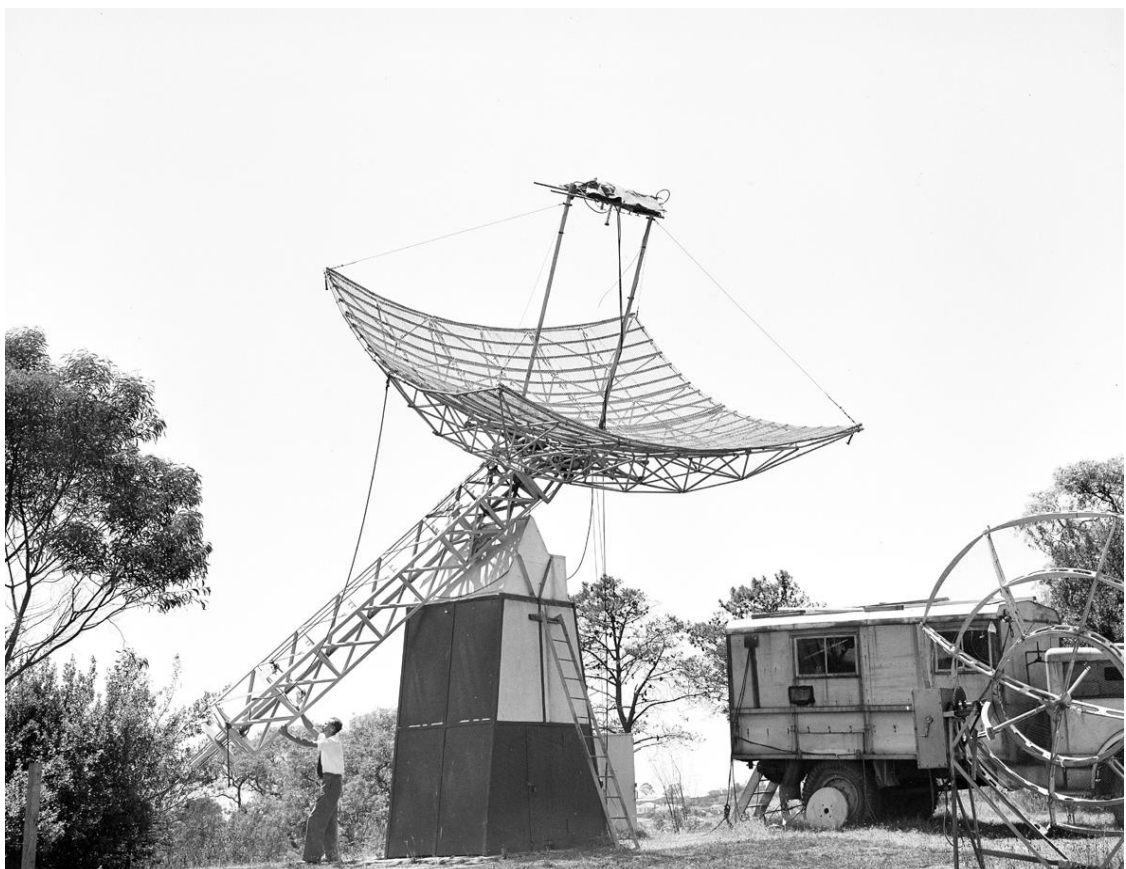

**Figure 26** - The 16-ft × 18-ft paraboloid at Potts Hill field station in January 1952.

In their published spectra Piddington and Minnett also included measurements obtained by Shain (1951) from the Hornsby Valley field station at 18.3 MHz and the 60, 100 and 160 MHz observations of Stanley and Slee (1950) from Dover Heights, giving a total of five frequency points on the radio spectrum. At this stage the contributions of the different components of Centaurus A (the inner lobes and the extended source) had not been recognised.

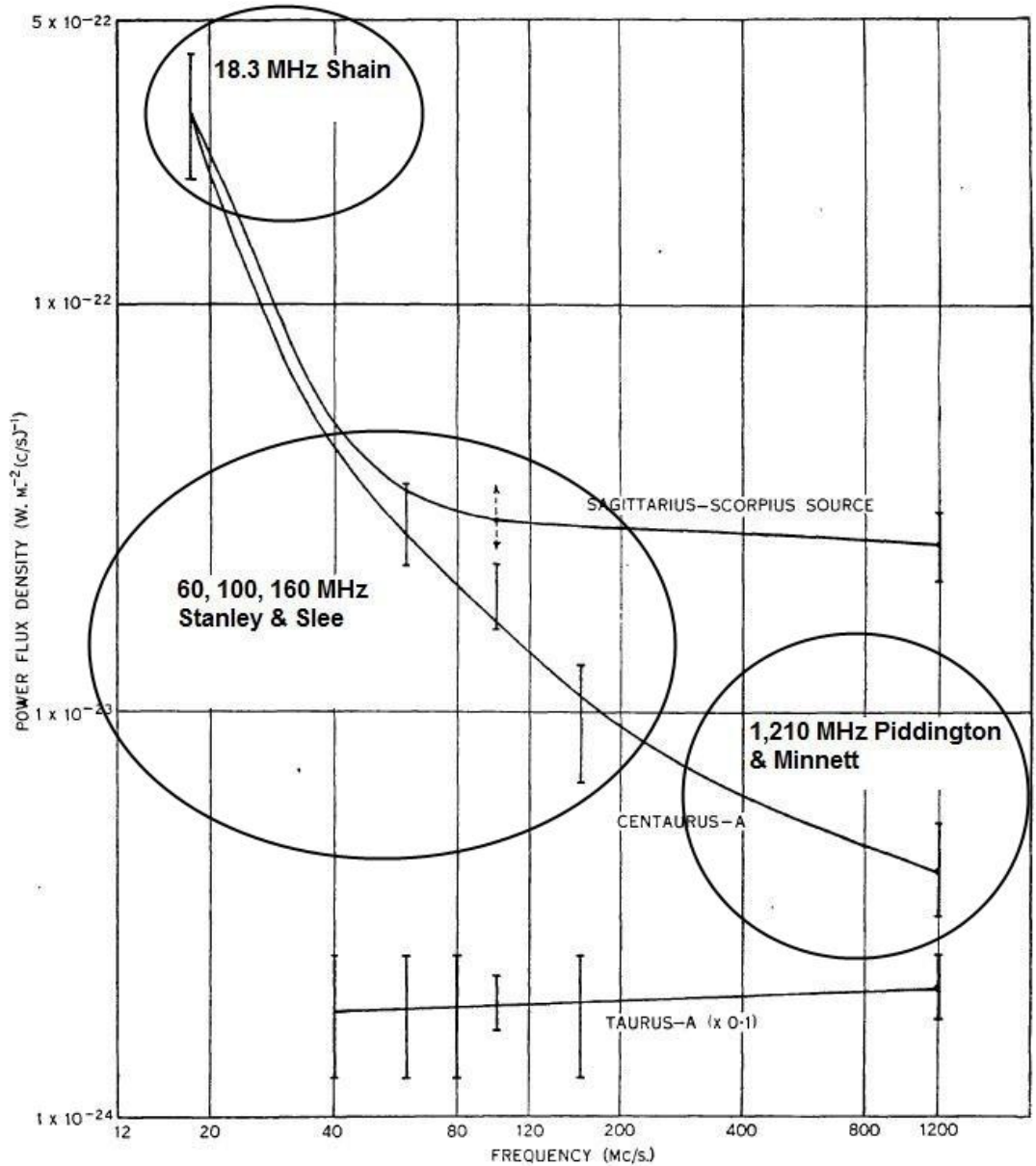

**Figure 27** - By 1951 the spectrum of Centaurus A measured by Radiophysics groups extended from 18.3 to 1210 MHz. The contributions of the flux density measurements from three groups are shown. The spectra of Sagittarius A and Taurus A are also shown. Note that the flux density of Taurus A has been reduced by a factor of 10 to separate it from the other spectra [after Piddington & Minnett 1951].

## 6.2 Hornsby Valley—1949–51

In May 1949, Alec Shain and Charles Higgins began observations using an array of eight half-wavelength dipoles arranged as a transit instrument located at Hornsby Valley field station (see Figure 28). Hornsby Valley was chosen for these observations as it was naturally shielded by the surrounding hills from the radio interference from the growing suburbs of Sydney. This first set of low-frequency observations continued until November 1949.

Following the initial success with observations at 18.3 MHz, Shain and Higgins decided to construct a new aerial at Hornsby Valley that would produce a higher resolution beam. The new aerial consisted of an array of 30 horizontal half-wave dipoles arranged in a grid suspended 0.2 wavelengths above the ground. This produced a beam of 17° between half-power points, considerably smaller than the ~45° beamwidth of the earlier 18.3 MHz system.

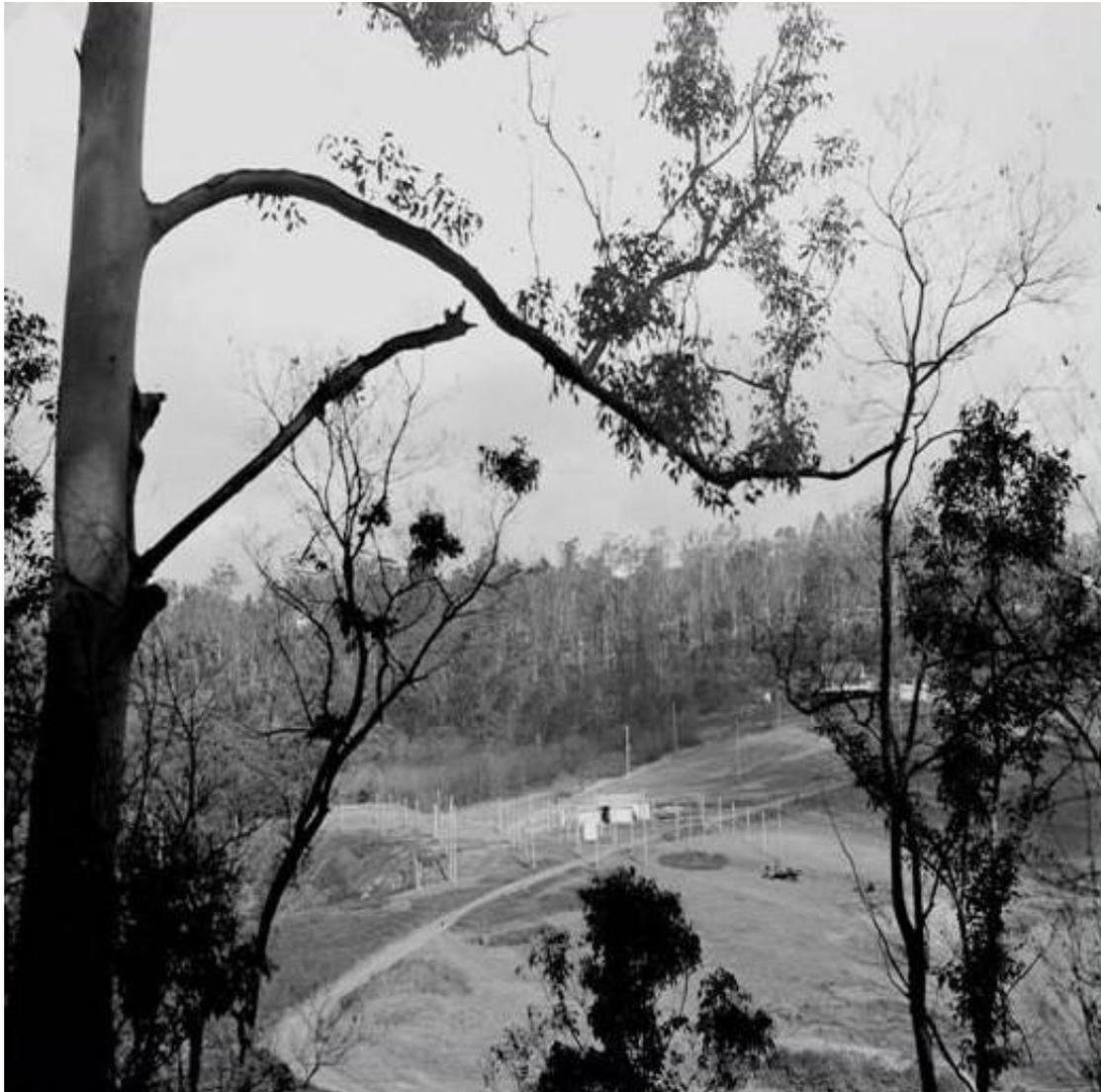

**Figure 28** - The Hornsby Valley field station in July 1952 showing the 18.3 MHz eight half-wave dipoles connected in phase and raised 0.2 wavelengths above the ground.

Observations using the new aerial system were made between June 1950 and June 1951, although the results were not published until 1954 (Shain & Higgins 1954). Based on their observations (see Figure 29), and after some fairly complex manipulations to extract the galactic component within the aerial beam, Shain and Higgins estimated the flux density of Centaurus A to be $5.3 \times 10^4$ Jy.

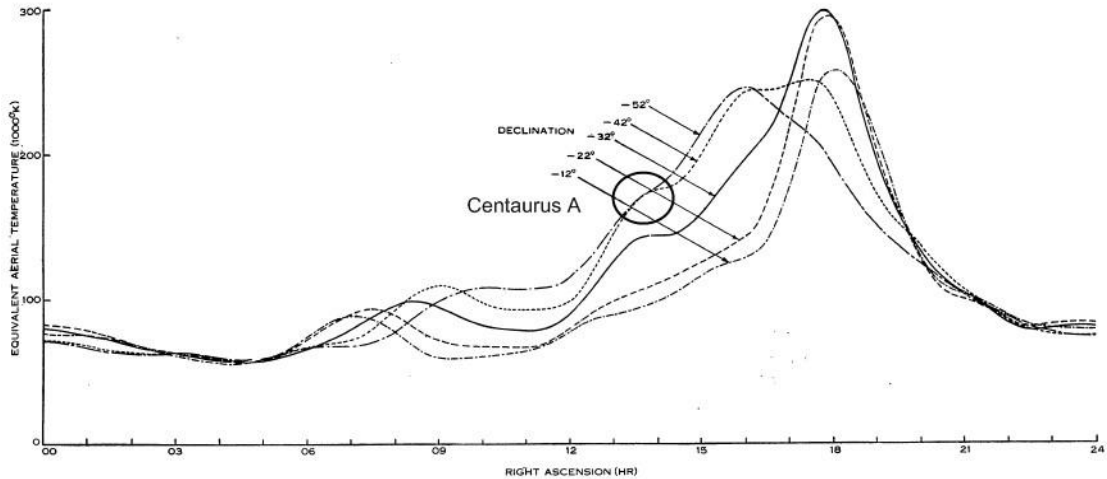

**Figure 29** - Observed equivalent aerial temperature as a function of RA for five different declinations observed using the 17° beam at 18.3 MHz. The declination scan for -42° is marked with the position of Centaurus A [after Shain & Higgins 1954].

## 6.3 Badgerys Creek—1950

Bernard Mills first observations in radio astronomy used interferometers at the Potts Hill field station. He initially borrowed time on Ruby Payne-Scott's 97 MHz swept lobe interferometer to try and determine an accurate position for Cygnus A (Mills & Thomas 1951). Encouraged by his initial experiences, Mills decided to build his own interferometer to further investigate the discrete radio sources. He chose a new independent site for these investigations at Badgerys Creek which was a CSIRO cattle research station. From February to December 1950, Mills (1952a) used a three element 101 MHz interferometer with a 60-m and 270-m baseline to obtain positions of 77 discrete sources including Centaurus A. The interferometer consisted of three broadside arrays (see Figure 30) arranged along an E–W baseline. Each of the broadside antennas could be rotated in elevation on its horizontal axis allowing a transit survey of the whole sky in different declination strips.

By simultaneously measuring the source at both interferometer spacings and by noting the ratio between the intensity of the interference fringes, Mills was able to obtain a rough estimate of the angular size of Centaurus A as ~20 arcmin in the east-west dimension and with a flux density of 1600 Jy at 101 MHz. At this fringe spacing Mills was detecting only the central source as the extended source lobes would have been fully resolved. Mills did note in his published results that larger intensities in the interference pattern were observed at the closest spacing ($D$ = 60 m, fringe spacing 2.8°), consistent with either an extended source in the E–W direction or alternatively two discrete sources of near equal intensity close together.

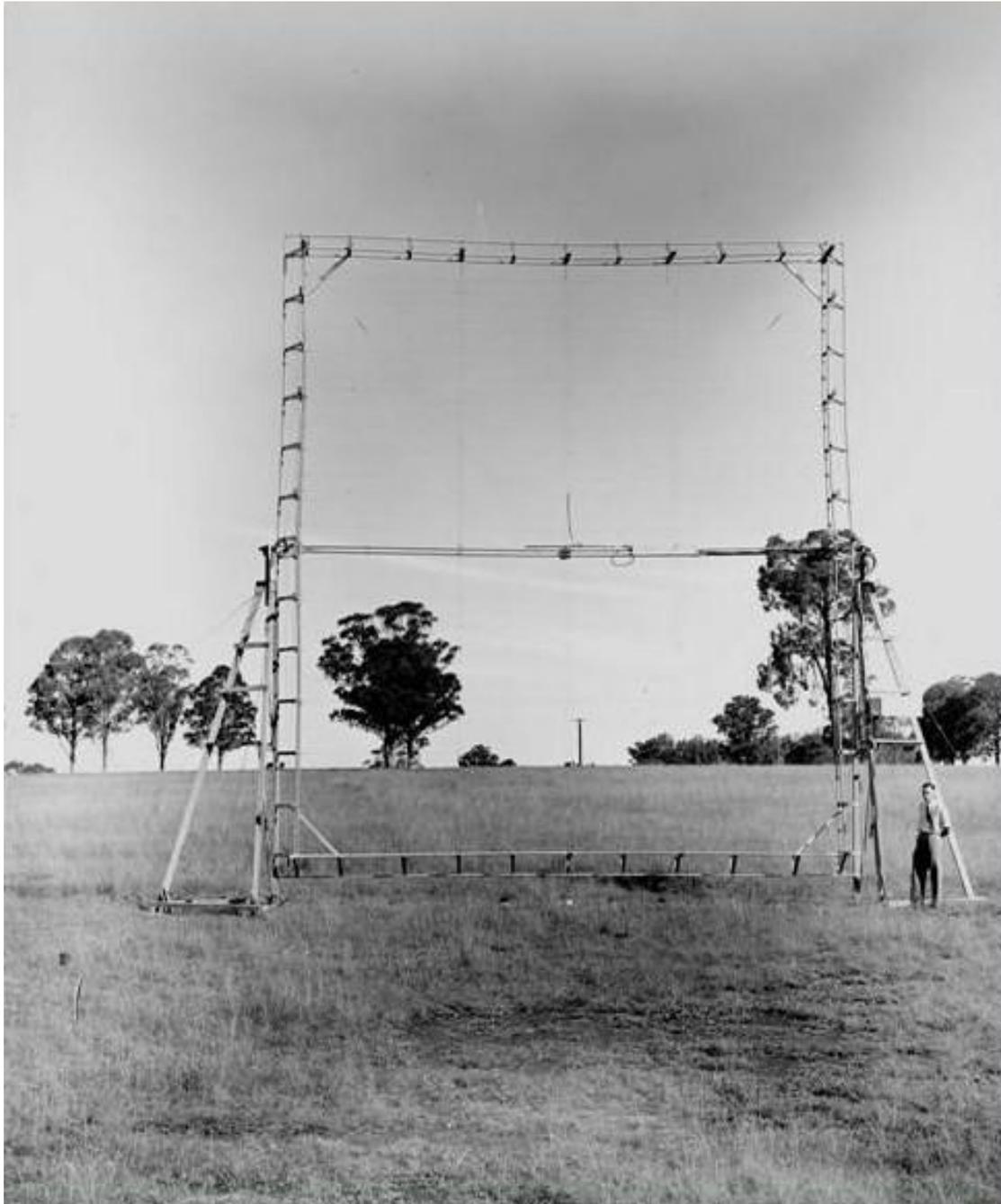

Figure 30 - One of the 101 MHz broadside antennas at Badgerys Creek field station in May 1950.

After this initial investigation Mills (1952b, 1952c) extended the baseline of his interferometer by introducing a radio link. This allowed him to obtain more accurate positions of six discrete radio sources and to achieve greater angular resolution of the sources in the E–W baseline. He found that Centaurus A at 101 MHz was fully resolved with the larger interferometer spacings and that the source had a strong central concentration with an E–W angular extent of 6 arcmin (the inner source lobes).

These observations also confirmed the earlier position correlation proposed by Bolton, Stanley & Slee (see Figure 31). Mills noted in his published results that Bolton (1953) had reported during the 1952 URSI meeting held in Sydney that his team's investigations at Dover Heights using an azimuth sea interferometer had found the source was nearly 2° in extent with a central concentration. The results of these observations were only published much later (Bolton et al. 1954). The existence of the two distinct components of the source was confirmed by Mills. He found at the closest interferometer spacing (60 m) that the extended component was of angular size ~2°, while the strong central concentration appeared to be only 6 arcmin in extent.

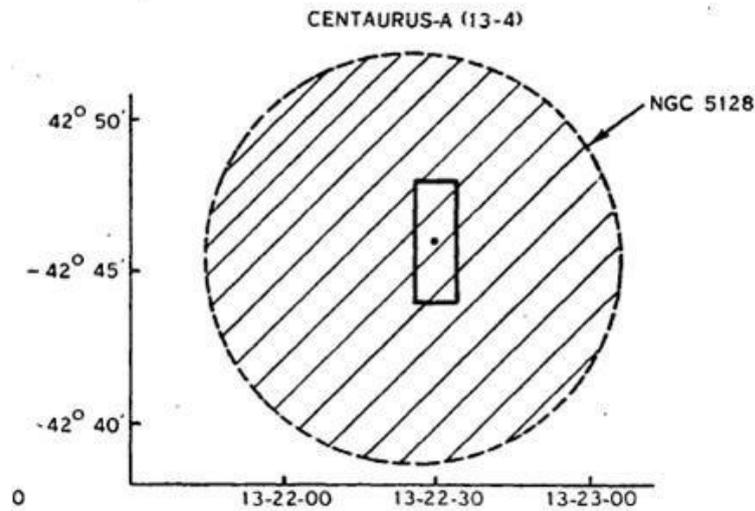

**Figure 31** - Position estimate for Centaurus A with error box obtained using a 101 MHz radio-linked interferometer. The optical extent of NGC 5128 is shown by the hash circle [after Mills 1952c].

Mills followed his initial observations with a more detailed investigation of radio-brightness distributions of four discrete sources including Centaurus A. Besides measuring the source distribution on an E–W baseline he also used two other baseline orientations to attempt to obtain a more comprehensive distribution. The measurements again revealed the complex nature of the brightness distribution. The undulating fit of the distribution (see Figure 32) shows that it is not a simple elliptical brightness distribution, but is likely due to the presence of the two inner lobes.

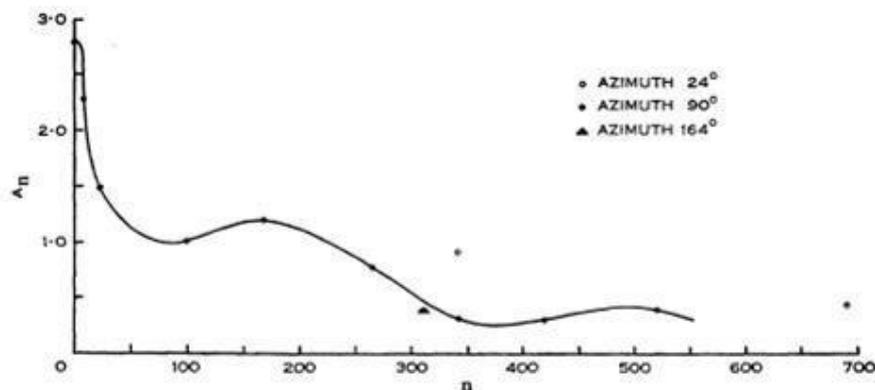

**Figure 32** - Mills' measurement of Centaurus A at 101 MHz showing the amplitude spacing spectra for three different baseline orientations, where *n* is the spacing in wavelengths [after Mills 1953].

Using the interferometer measurements, Mills constructed an equivalent radio-brightness distribution image of the central concentration of the radio source (see Figure 33). While noting that it was difficult to obtain an accurate distribution with the limited number of spacings and baseline orientations, Mills proposed that the central component of the source has an elliptical distribution with an orientation similar to that of the dust-band visible in the optical image. He concluded that this may suggest that the radio emission was associated with gas in this region, although he noted that the radio emission appeared to be non-thermal. It is likely that the use of the simplifying symmetry assumption as well as the limited baseline orientations distorted the estimated source distribution. Comparison with a high resolution measurement of the inner radio lobes shows they are oriented at 90° to that estimated by Mills.

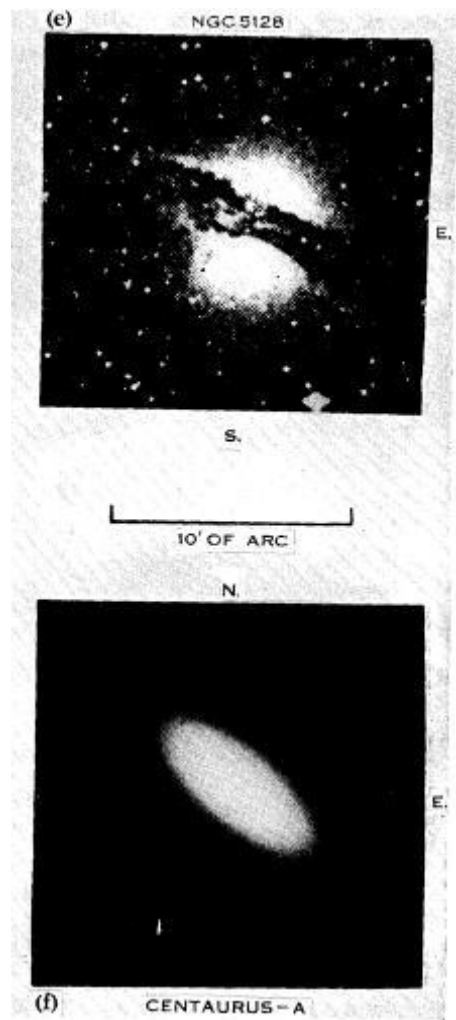

**Figure 33** - Comparison of an optical image of Centaurus A (above) with a radio-brightness distribution image (below) which was constructed based on the 101 MHz interferometer measurements [after Mills 1953].

### 6.4 Potts Hill—1955

In 1955, Jack Piddington and Gil Trent used the 36-ft transit parabola at Potts Hill (see Figure 34) to conduct a 600 MHz survey of the southern sky. The 36-ft transit aerial had been constructed for a dedicated hydrogen-line survey by a team led by Frank Kerr, following the discovery of the 21-cm line in 1951. At 600 MHz the 36-ft aerial produced a 3.3° beamwidth and Piddington and Trent (1956a, 1956b) were able to conduct a survey between declinations of 90° S and 50° N. From this survey they identified 49 discrete sources with a flux density greater than 100 Jy, including Centaurus A.

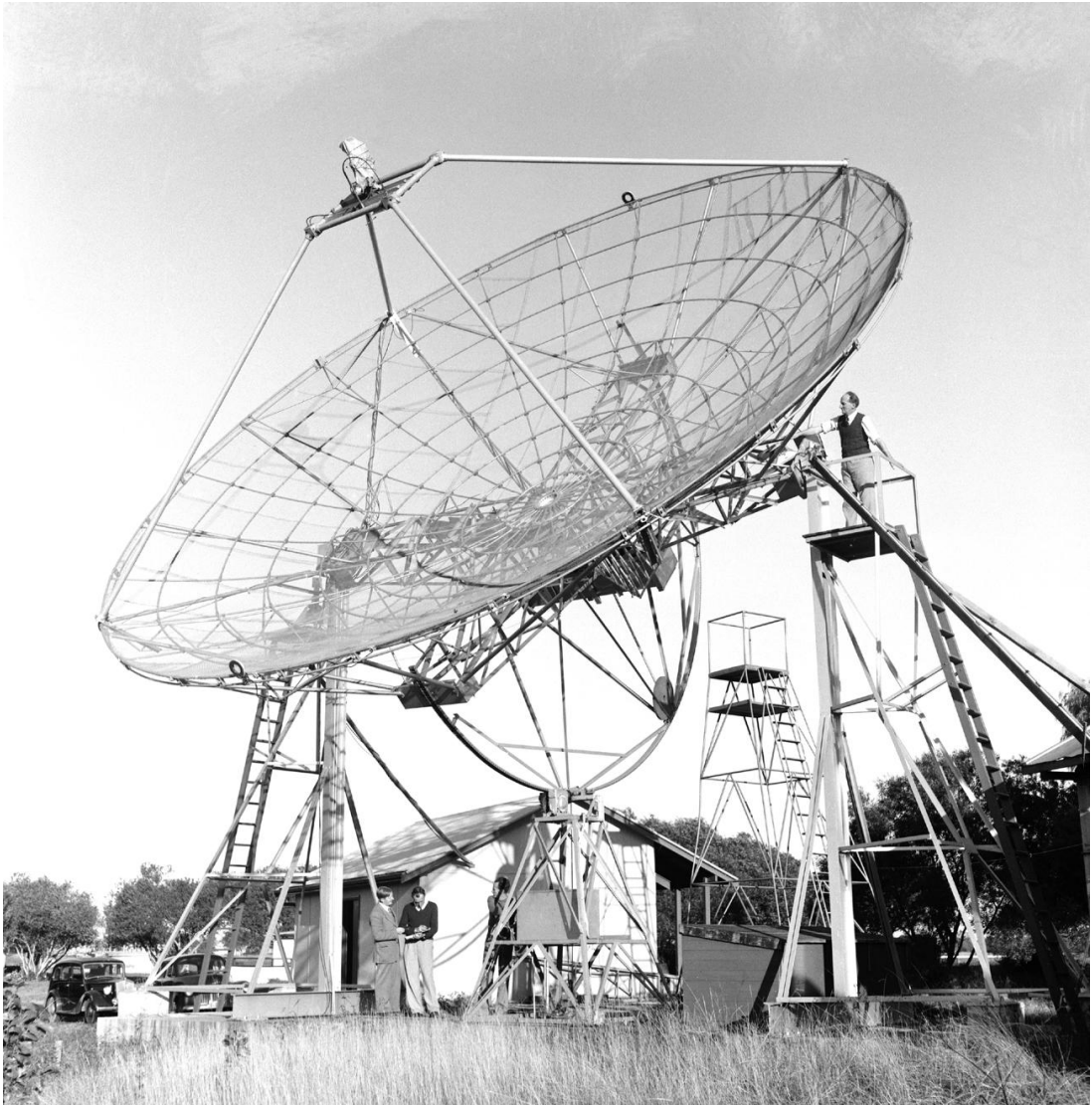

**Figure 34** - The 36-ft transit parabola at Potts Hill in June 1955.

Piddington and Trent noted that Centaurus A consisted of an extended source ~2° in diameter (see Figure 35), concentric with the more concentrated central source that had been discovered by Bolton's group. At 600 MHz the total flux density of the source was determined to be 1300 Jy. Piddington and Trent also noted that the source appeared to overlap the background emission of a narrow region which joined the main galactic emission (see Figure 36).

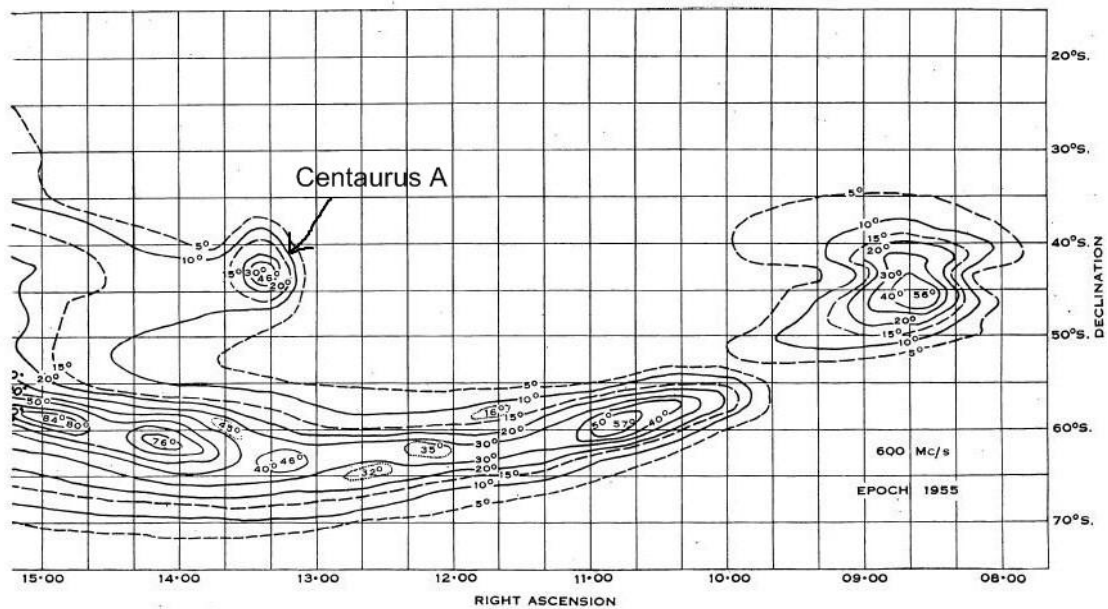

**Figure 35** - Radio isophotes of 600 MHz continuum emission from southern sky survey, showing the section containing Centaurus A. The large angular extent of Centaurus A is clearly visible [after Piddington & Trent 1956b].

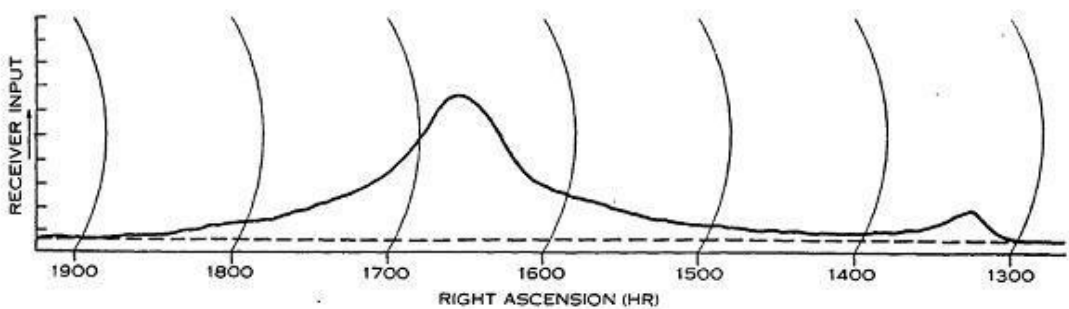

**Figure 36** - Individual 600 MHz record taken at declination 44°.9 S and between RA 19:00 and 13:00. The strong source at RA 16:40 is the Galactic Plane. The extension of the background emission extending to Centaurus A at RA 13:20 is evident. The graduation of the receiver input temperature is ~8 K [after Piddington & Trent 1956a].

### 6.5 Fleurs—1954–58

After experimenting with interferometer designs at Badgerys Creek, Mills devised a new instrument that became known as the Mills Cross. This design combined the advantages of an interferometer, allowing a large aperture (and hence high resolution) to be achieved at low cost, while having the advantage of a pencil-beam instrument in being able to directly measure a brightness distribution without the need for a complex Fourier transformation. Mills and Little first tested a prototype of the new instrument at Potts Hill before a full scale 85.5 MHz instrument was constructed at the new Fleurs field station during 1953–54. The full scale Mills Cross produced a pencil-beam response with a beamwidth of 50 arcmin (see Figure 37).

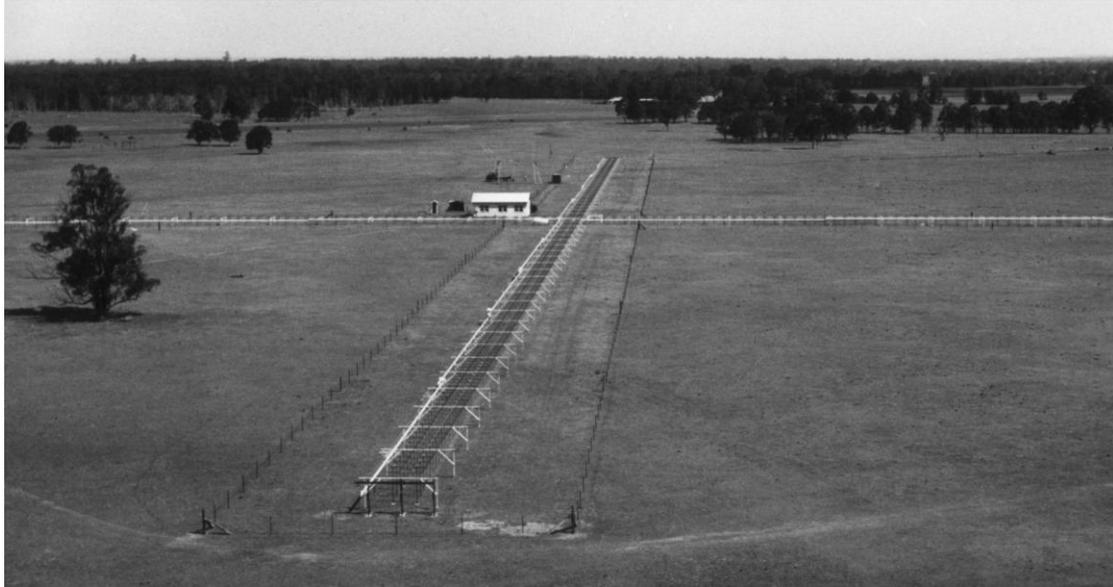

**Figure 37** - The Mills Cross at the Fleurs field station in October 1954 looking south along the N-S arm.

One of the early detailed observations made using the Mills Cross was of Centaurus A (Sheridan 1958). While the 50 arcmin beam was unable to fully resolve the central concentration of the source, the full extent of the extended source was clearly revealed (see Figure 38) and was much larger than the earlier estimates of ~2°.

Sheridan (1958) noted that, assuming a distance to NGC 5128 of ~750 kpc, the linear extent of the radio source at 85.5 MHz would be 90 kpc by 30 kpc. More recent distance estimates place Centaurus A at 3.4±0.15 Mpc (Israel 1998) , although this still makes Centaurus A by far the closest active radio galaxy. At this distance, 1 arcmin of the sky corresponds almost precisely to 1 kpc, revealing the truly massive size of the extended radio lobes. At this increased distance the power emitted and its linear size are increased by factors of 20.3 and 4.5 respectively.

A second cross-instrument, know as the Shain Cross, was constructed in 1956 at Fleurs field station operating at 19.7 MHz with a beamwidth of 1.4° (see Figure 39). Shain (1958) used this instrument to obtain isophotes of Centaurus A at 19.7 MHz which showed the radio source extended approximately 7.5° in a north–south direction and 2.5° in the east–west direction (see Figure 40). This was slightly larger again than that observed at 85.5 MHz.

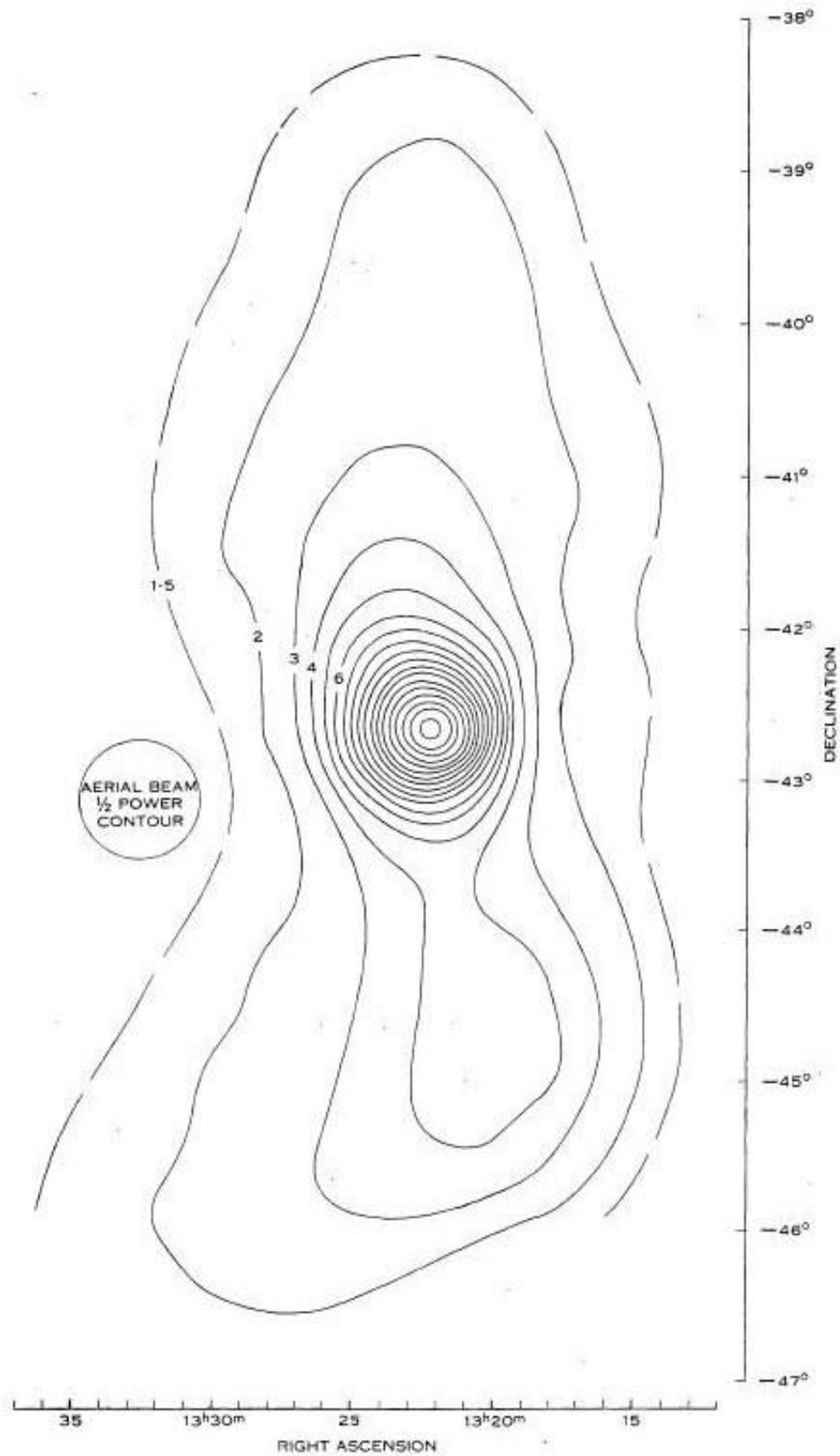

**Figure 38** - Radio isophotes of Centaurus A at 85.5 MHz observed by the Mills Cross at the Fleurs field station.  The contour interval is 3000 K [after Sheridan 1958].

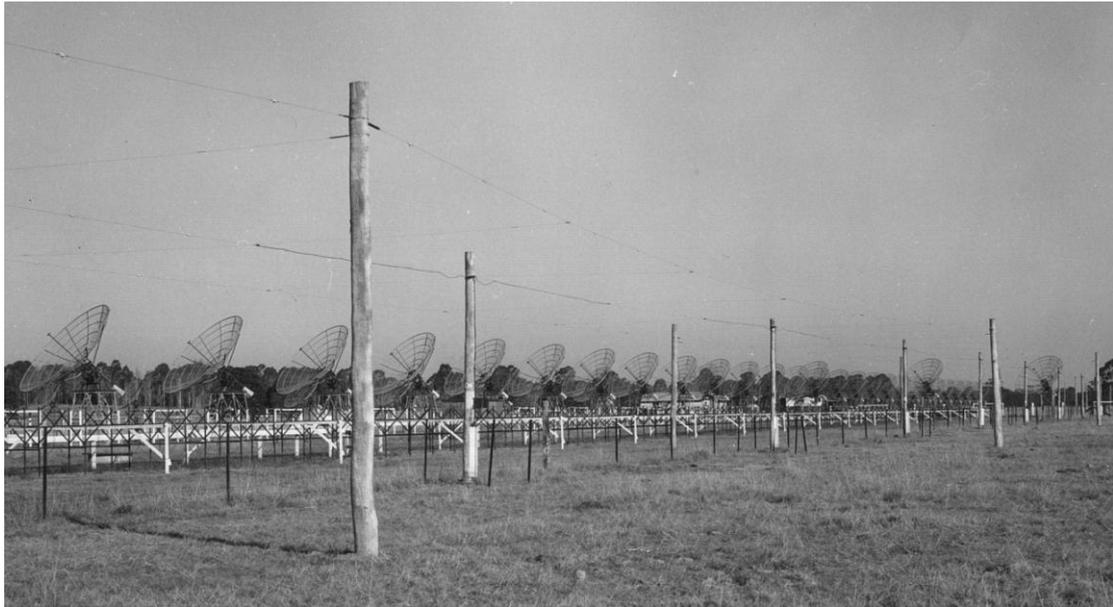

**Figure 39** - Suspended dipoles of the Shain Cross at the Fleurs field station. In the background are the aerials of the Chris Cross.

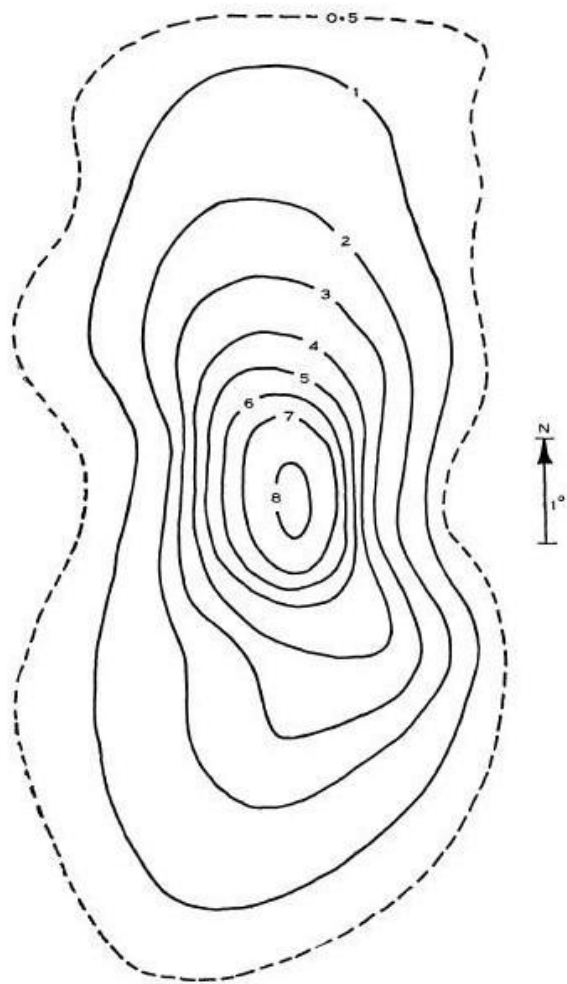

**Figure 40** - Isophotes of Centaurus A as observed at 19.7 MHz using the Shain Cross at the Fleurs field station [after Shain 1958].

Shain noted that at 19.7 MHz, 11% of the total flux was received from the central source, while at 85.5 MHz the proportion is 23% (Sheridan 1958). Based on these observations, Shain estimated the spectra of the two separate source components using the earlier Radiophysics observations (see Figure 41). The Stanley & Slee (1950) observations at 60, 100 and 160 MHz and some of the Mills (1952c) 101 MHz observations were made with closely spaced interferometer fringes which would only have detected the central source. The Piddington & Minnett (1951) observations at 1210 MHz were made with an aerial beamwidth of 2.8° and as such would have accepted only a small fraction of the extended source, which at the higher frequency, was also much weaker. Piddington & Trent (1956a) measured the integrated flux density from the whole source at 600 MHz, while McGee, Slee & Stanley (1955) obtained contours at 400 MHz from which the total flux density could be derived. All of these observations were used to construct the spectra shown in Figure 41.

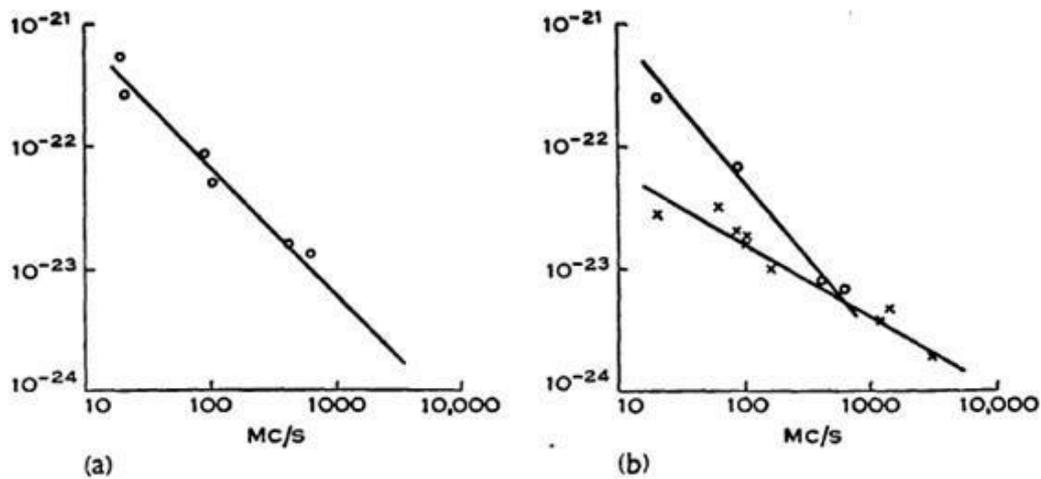

**Figure 41** - Shain (1958) obtained updated spectra for Centaurus A based on measurements at 19.7 MHz and earlier Radiophysics measurements: (a) The spectrum of the total flux density of Centaurus A with a slope of -1.0. (b) The spectra of the two different components of Centaurus A. The extended source has a slope of -1.25, while the central source has a slope of -0.6. The circles are flux measurements from the extended source, while the crosses are from the central source [after Shain 1958].

Based on the updated spectra, Shain estimated that the total received flux between 10 and 3000 MHz was $3.9 \times 10^{-14}$ W m$^{-2}$. Using the distance to NGC 5128 of 3.4 Mpc quoted by Israel (1998), the radiated power at 19.7 MHz from the whole source is $5.5 \times 10^{33}$ W and its linear size is $400 \times 140$ kpc.

### 6.6 Potts Hill—1958

During 1958, Jim Hindman and Cam Wade used the 36-ft transit aerial to observe Centaurus A, but in this instance at 1400 MHz using Kerr's 21-cm receiver modified for reception of continuum radiation. At this frequency the beamwidth was 1.4°. They found that the source extended ~7° in declination and ~3° in RA with a similar distribution to that observed by Sheridan at 85.5 MHz. Hindman & Wade (1959) found the total flux density of the source to be $1.3 \times 10^4$ Jy and that 23% of this could be attributed to the central source (see Figure 42).

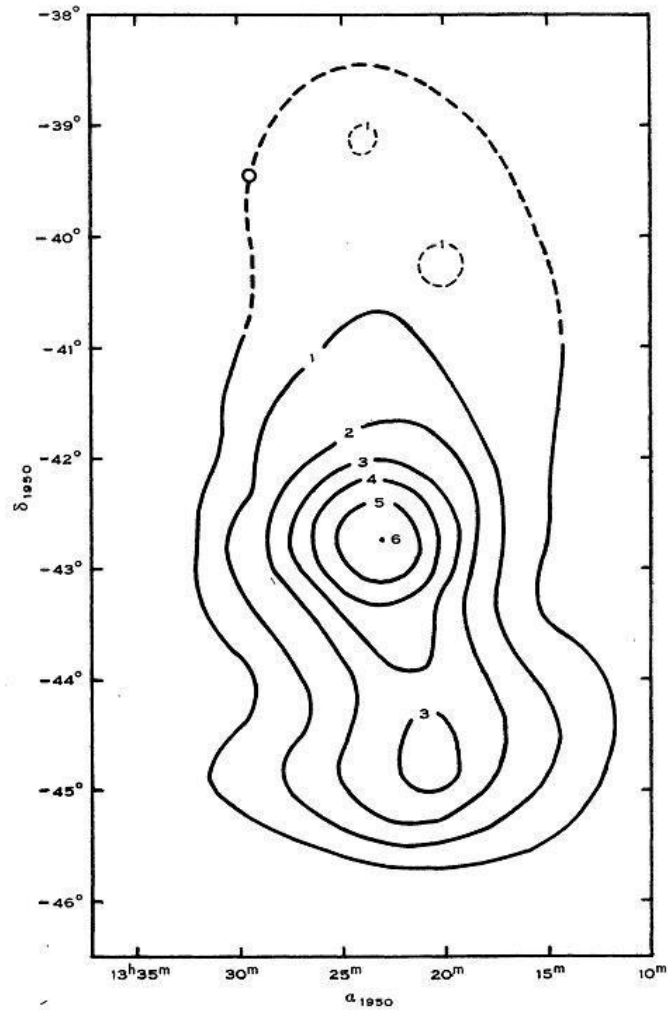

**Figure 42** - Isophotes of equivalent aerial temperature at 1400 MHz [after Hindman & Wade 1959].

In a more detailed paper examining the extended component of Centaurus A, Wade (1959) used both 1400 and 85.5 MHz data to subtract the contribution of the central source from the extended source to reveal, for the first time, the double lobe nature of the extended source, which is now commonly associated with active radio galaxies (see Figure 43).

## 7  Conclusions

The year 1961 marked a watershed in the development of radio astronomy in Australia.  Apart from the groups studying solar radio astronomy, most of the Radiophysics field stations in the Sydney area were shut down in preparation for the opening of a large parabolic dish under construction near the township of Parkes, 250 km west of Sydney.  Although smaller than the first large dish built at Jodrell Bank in England (64 m diameter compared with 76 m), the Parkes dish was technically a much superior instrument (see e.g. Robertson 1992).

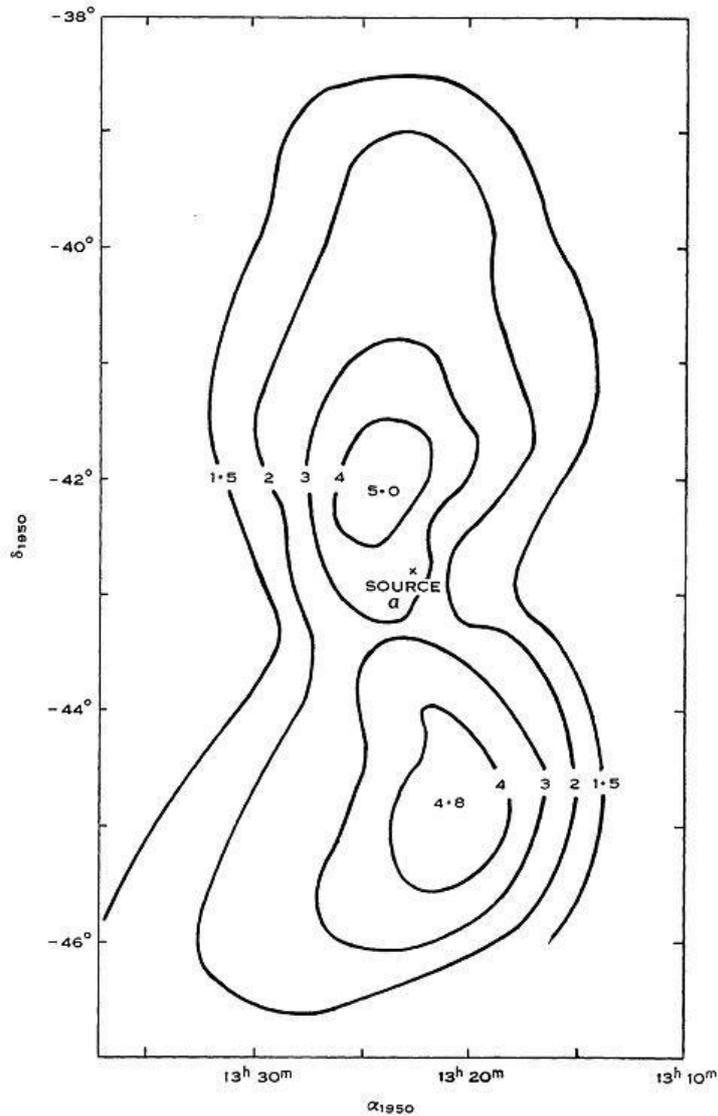

**Figure 43** - Isophotes at 85.5 MHz of Centaurus A after subtracting radiation due to the central source which is marked as 'source a' in the diagram. This clearly shows for the first time the double nature of the extended source [after Wade 1959].

Within months of the inauguration of the telescope in October 1961, Radiophysics staff at the Parkes Observatory made two discoveries of major significance. Both discoveries were made by observations of Centaurus A. The first was that radio emission from Centaurus A shows a significant degree of polarisation, the most convincing evidence until then that radio sources produce their prodigious amounts of radio energy by the synchrotron process (see Bracewell, Cooper, & Cousins 1962; Gardner & Whiteoak 1962). The second and even more surprising result was that the plane of polarisation undergoes Faraday rotation, where the amount of rotation depends on the wavelength of the radiation, and thus opened a way of measuring the magnetic fields of our galaxy (see Cooper & Price 1962). Both discoveries quickly established the Parkes Observatory at the forefront of radio astronomy, a position it has maintained to the present day.

When James Dunlop first observed NGC 5128 on the night of 29 April 1826 at the Parramatta Observatory, and mistakenly described it as two separate nebulae lying side by side, he could not have foreseen the central role this nebula – Centaurus A – would play in the development of Australian astronomy. He was the first astronomer to observe the southern skies with a telescope of significant power and he discovered over 320 southern non-stellar objects. While Charles Messier in Paris was renowned for cataloguing the brightest 'deep-sky' objects visible from the northern hemisphere, James Dunlop can be rightly called the 'Messier of the Southern Sky'.

## Acknowledgments

We are grateful to the Historical Photographic Archive at the Australia Telescope National Facility; all images are courtesy of this archive unless indicated otherwise. We also acknowledge the assistance of staff at the NSW branch of the National Archives of Australia at Chester Hill in Sydney. One of us (PR) acknowledges travel support from CSIRO to present a shorter version this paper at the International Conference on the Many Faces of Centaurus A in Sydney in July 2009.

## References

Bergman, G. F. J., 1967, Rümker, Christian Carl Ludwig (1788–1862), Australian Dictionary of Biography, vol. 2 (Melbourne: Melbourne University Press), 403
Bolton, J. G., 1948a, letter to E. G. Bowen 15 June 1948 (National Archives of Australia)
Bolton, J. G., 1948b, Nature, 162, 141
Bolton, J. G., 1948c, New Zealand Cosmic Noise Expedition and Dover Heights Report No. 5: Position of Centaurus A, Radiophysics Laboratory (National Archives of Australia)
Bolton, J. G., 1953, Observatory, 73, 26
Bolton, J. G., 1982, PASA, 4, 349
Bolton, J. G. & Slee, O. B., 1953, AuJPh, 6, 420
Bolton, J. G. & Stanley, G. J., 1948, Nature, 161, 312
Bolton, J. G., Stanley, G. J. & Slee, O. B., 1949, Nature, 164, 101
Bolton, J. G., Stanley, G. J. & Slee, O. B., 1954, AuJPh, 7, 110
Bolton, J. G., Westfold, K. C., Stanley, G. J. & Slee, O. B., 1954, AuJPh, 7, 96
Bowen, E. G., 1948, letter to the CSIR Executive, undated (National Archives of Australia)
Bracewell, R. N., Cooper, B. F. C. & Cousins, T. E., 1962, Nature, 195, 1289
Brisbane, T. M., 1835, Catalogue of 7385 Stars, Chiefly in the Southern Hemisphere (London: Lords Commissioners of the Admiralty)
Burns, J. O., Feigelson, E. D. & Schreier, E. J., 1983, ApJ, 273, 128
Cooper, B. F. C. & Price, R. M., 1962, Nature, 195, 1084
Dunlop, J., 1828, Phil Trans Roy Soc, 118, 113
Dunlop, J., 1829, Mem Roy Astron Soc London, 3, 257
Gardner, F. F. & Whiteoak, J. B., 1962, PhRev, 9, 197
Haynes, R., Haynes, R., Malin, D. & McGee, R. X., 1996, Explorers of the Southern Sky (Sydney: Cambridge Univ Press)
Herschel, J. F. W., 1848, Results of Astronomical Observations made during the years 1834, 5, 6, 7, 8 at the Cape of Good Hope, being a completion of a telescopic survey of the whole surface of the visible heavens commenced in 1825 (London: Smith, Elder & Co.)
Hey, J. S., 1946, Nature, 157, 47
Hey, J. S., Parsons, S. J. & Phillips, J. W., 1946, Nature, 158, 234
Heydon, J. D., 1966, Brisbane, Sir Thomas Makdougall (1773–1860), Australian Dictionary of Biography, vol. 1 (Melbourne: Melbourne University Press), 151
Hindman, J. V. & Wade, C. M., 1959, AuJPh, 12, 258
Israel, F. P., 1998, A&ARv, 8, 237
Lomb, N., 2004, Hist Rec Aust Sci, 15, 211

McCready, L. L., Pawsey, J. L. & Payne-Scott, R., 1947, Proc. R. Soc. London, A190, 357

McGee, R. X. & Bolton, J. G., 1954, Nature, 173, 985

McGee, R. X., Slee, O. B. & Stanley, G. J., 1955, AuJPh, 8, 347

Mills, B. Y., 1952a, AuJSciRes, 5, 266

Mills, B. Y., 1952b, Nature, 170, 1063

Mills, B. Y., 1952c, AuJSciRes, 5, 245

Mills, B. Y., 1953, AuJPh, 6, 452

Mills, B. Y. & Thomas, A. B., 1951, AuJSciRes, 4, 158

Morrison-Low, A. D., 2004, Hist Rec Aust Sci, 15, 151

Norris, R., 2008, Austral Sci 29(4), 16

Orchiston, W., 1994, AuJPh, 47, 541

Orchiston, W., 2005, Dr Elizabeth Alexander: first female radio astronomer, In The New Astronomy: Opening the Electromagnetic Window and Expanding our View of Planet Earth (New York: Springer), 71

Orchiston, W. & Slee, O. B., 2002a, AAO Newsl., 101, 25

Orchiston, W. & Slee, O. B., 2002b, J Astr Hist Heritage, 5, 21

Orchiston, W. & Slee, O. B., 2005, The Radiophysics field stations and the early development of radio astronomy, In The New Astronomy: Opening the Electromagnetic Window and Expanding our View of Planet Earth (New York: Springer), 119

Payne-Scott, R., Yabsley, D. E. & Bolton, J. G., 1947, Nature, 160, 256

Piddington, J. H. & Minnett, H. C., 1951, AuJSciRes, 4, 459

Piddington, J. H. & Trent, G. H., 1956a, AuJPh, 9, 74

Piddington, J. H. & Trent, G. H., 1956b, AuJPh, 9, 481

Rivett, C., 1988, Parramatta Bicentenary: Australia (Parramatta: Richard Pike)

Robertson, P., 1992, Beyond Southern Skies: Radio Astronomy and the Parkes Telescope (Sydney: Cambridge Univ. Press)

Saunders, C. D., 1990, Astronomy in Colonial New South Wales: 1788 to 1858, Sydney, 151

Saunders, S. D., 2004, Hist Rec Aust Sci, 15, 177

Service, J., 1890, Thir Notandums, Being the Literary Recreations of Laird Canticarl of Mongrynen: To Which is Appended a Biographical Sketch of James Dunlop, Edinburgh, 135

Shain, C. A., 1951, AuJSciRes, 4, 258

Shain, C. A., 1958, AuJPh, 11, 517

Shain, C. A. & Higgins, C. S., 1954, AuJPh, 7, 130

Sheridan, K. V., 1958, AuJPh, 11, 400

Slee, O. B., 1955, AuJPh, 8, 498

Slee, O. B., 1977, AuJPh, Astrophys Suppl No. 43

Stanley, G. J. & Slee, B., 1950, AuJSciRes, 3, 234

Wade, C. M., 1959, AuJPh , 12, 471

Wood, H., 1966, Dunlop, James (1793–1848), Australian Dictionary of Biography, vol. 1 (Melbourne: Melbourne University Press), 338